\documentclass{webofc}
\usepackage[varg]{txfonts}

\graphicspath{{Fig/}}

\newcommand{\grad}{\bf \nabla}

\newcommand{\derive}[2]{\frac{\partial #1}{\partial #2}}
\newcommand{\derived}[2]{\frac{\partial^2 #1}{\partial #2^2}}

\newcommand{\beqa}{\begin{eqnarray}}
\newcommand{\eeqa}{\end{eqnarray}}

\newcommand{\nab}{ {\bf \nabla} }
\newcommand{\lapl}{ \Delta }

\newcommand{\beq}{\begin{equation}}
\newcommand{\eeq}{\end{equation}}

\newcommand{\greq}{\begin{equation} \begin{array}{l}}
\newcommand{\egreq}{\end{array} \end{equation}}

\newcommand{\eeqn}[1]{\label{#1}\end{equation}}

\newcommand{\beqan}{\begin{eqnarray}}
\newcommand{\eeqan}[1]{\label{#1}\end{eqnarray}}

\newcommand{\cms}{$\mathrm {cm}^2 {s}^{-1}$}

\begin{document}
\title{Turbulence in stably stratified radiative zone}
%
%

\author{\firstname{François} \lastname{Lignières}\inst{1}\fnsep\thanks{\email{francois.lignieres@irap.omp.eu}}}

\institute{Institut de Recherche en Astrophysique et Plan\'etologie, Université de Toulouse, CNRS, CNES, UPS, France
          }

\abstract{%
The topic of turbulent transport in a stellar radiative zone is vast and poorly understood. 
Many physical processes can potentially drive turbulence in stellar radiative zone but the limited observational 
constraints and the uncertainties in modelling turbulent transport make it difficult to identify the most relevant one.
Here, we focus on the effect of stable stratification on the radial turbulent transport of chemicals and more particularly on the case
where the turbulence
is driven by a radial shear. Results of numerical simulations designed to test phenomenological models of turbulent transport will
be presented.
While this may appears little ambitious, stable stratification will influence
the radial turbulent transport whatever the mechanism that drives the turbulent motions and the present considerations
should be useful for these other mechanism too.
}
\maketitle
\section{Introduction}
\label{intro}

Most constraints on the transport of angular momentum and chemicals in radiative zones are indirect. Maybe, the most direct one is
the observation of abundance anomalies at the surface of intermediate-mass stars, called chemically peculiar stars. 
These anomalies are due to the gravitational settling and radiative levitation of chemical elements, both being
slow processes. This can only happen if the stably stratified subphotospheric layers are nearly quiescent showing
that the macroscopic hydrodynamic transport is very inefficient there \cite{Michaud04}.
In other stars surface abundances are the signature
of deep mixing instead. This is the case in massive stars where the observed
surface abundances of elements involved in the CNO cycle require mixing down to the stellar core (e.g. \cite{Hunter2009}) or
in the Sun where the surface Lithium depletion indirectly shows that a radial transport occurs in the radiative zone below the convective envelope
(e.g. \cite{Pinsonneault1997}).
Then, helio and asteroseismology provide detailed informations on the interior of certain stars, 
in particular the sound speed, the Brunt-Väisälä frequency and the rotation rate, the latter being the best data available
to understand the dynamics \cite{Thompson1996, Mosser2012, Deheuvels2014, VanReeth2016}. 
All these constraints are compared with results of 1D stellar evolution codes that include the transport of chemical elements and angular momentum
using radial diffusion models (e.g. \cite{Salaris2017}).
This comparison provides estimates of the transport required to reproduce the surface and/or seismic data.

Contrary to planetary atmospheres or stellar convective envelopes, we have no direct constraint on the length scale and the velocity
that characterize the flow in radiative zones.
We know however one important thing, that is the radial angular transport, whatever its cause, is weak in the sense that
rotation velocities $\sim 2-100 \times 10^5 \mathrm{cm s}^{-1}$ and lengthscales $\sim 10^{11} \mathrm{cm}$ only produce an effective radial transport
with diffusion coefficient $D_{\rm eff}$ of the order
$10^3-10^4$ \cms in solar type stars \cite{Michaud1998, Korn2006, Eggenberger2012} and up to $10^8$ \cms in the external layers of massive star models
\cite{Salaris2017}.
In a radiative zone, the heat transport is ensured by radiation and the associated thermal diffusivity, $\kappa$, of the order of
$10^7$ \cms in the solar radiative zone, is much higher than $D_{\rm eff}$.
Thus the motions that are responsible for the effective transport of chemicals or angular momentum should
generate a heat transport that is negligible with respect to the radiative heat transport.
This means that, contrary to typical planetary atmospheres, 
the mean radial thermal stratification is not expected to be modified by hydrodynamical transport. This assumption is implicit in stellar evolution codes.

The most obvious reason for the small radial transport in radiative zones is the strength of the stable stratification relative to the dynamics.
The Brunt-Väisälä frequency $N$, that measures the time scale of the restoring buoyancy force $1/N$, is of the order of  a 
fraction of the star dynamical frequency $(GM/R^3)^{1/2}$. Except may be for the most rapid rotators, it will dominate motions taking place
on a rotation time scale $1/\Omega$ or
those induced by differential rotation if the gradient scale if of the order of the radius $r \frac{d \Omega}{d r} \sim \Omega$.
The $N/\Omega$ ratio is $\sim 360$ in the solar radiative zone and is still $\sim 15$ in a main-sequence intermediate-mass star
with a rotation period of $1$ day.

While radial motions are strongly limited by the buoyancy force, horizontal motions are not and may drive
an efficient transport in these directions.
Hydrodynamical models of chemical and angular momentum transport in stellar radiative zones have been developed under the assumption
that the horizontal transport is efficient enough to strongly limit the differential rotation in latitude.
Distinguishing a large scale axisymmetric meridional circulation from 3D turbulent motions, Zahn \cite{Zahn1992} derived
radial equations for the transport of angular momentum and chemical elements where the circulation velocities and the turbulent diffusion coefficients 
are expressed in terms of the prognostic variables $\Omega(r), C(r) ..$.
When compared to observations, this model has been quite successful for massive stars where a dominant transport process is the
radial transport induced by radial differential rotation \cite{Meynet2000, Maeder2001}. It is this process that we shall consider in detail in this lecture. 
However, for solar-type stars, the model predicts larger rotation rates than observed in the interior of the Sun \cite{Talon2005} and of the sub-giants \cite{Eggenberger2012}.
This led to consider gravity waves generated by convective motions at the radiative/convective interface as a possible
alternative for transport in solar-type stars \cite{Zahn2013}. Instabilities involving the magnetic fields are also considered as potential candidates to increase 
the angular momentum transport \cite{Spruit2002}.
We shall not consider these processes in the following, although as long as they involve
turbulent motions they will be also affected by stable stratification. 

In the following, we first consider the shear instability of parallel flow in stellar radiative zones (Sect.~\ref{shear}) and
then
discuss models of the vertical turbulent transport in stably stratified turbulence (Sect.~\ref{turb}). 
Recent numerical simulations allowed us to test models for the vertical transport of chemicals driven by a vertical shear flow in stellar conditions
and we shall 
present their results.

\section{Stability of parallel shear flows in radiative atmosphere}
\label{shear}

The stability of parallel shear flows is first considered in a fluid of constant density (Sect.~\ref{sec-shear}), then in the
presence of a vertical
stable stratification (Sect.~\ref{sec-strat}) and finally adding the effect of a high thermal diffusivity as in stellar interiors (Sect.~\ref{lign99}).

\subsection{Unstratified shear flows}
\label{sec-shear}

Let us first consider the kinetic energy potentially available in shear flows.
If we consider an inviscid  parallel flow ${\bf U} = U(z) {\bf e_x}$ between two horizontal plates where the vertical velocity vanishes, the initial 
horizontal momentum integrated across the layer $\int_0^H {\bf U}  dz$ is conserved.
The 
uniform flow $U_m {\bf e_x}$ that has the same total horizontal momentum, determined by $U_m H = \int_0^H U dz$ has a lower kinetic energy than the initial flow. 
Shear instabilities can then be viewed as
a mechanism to extract the kinetic energy stored in the shear.

\begin{figure*}
\centering
\resizebox{\hsize}{!}{\includegraphics{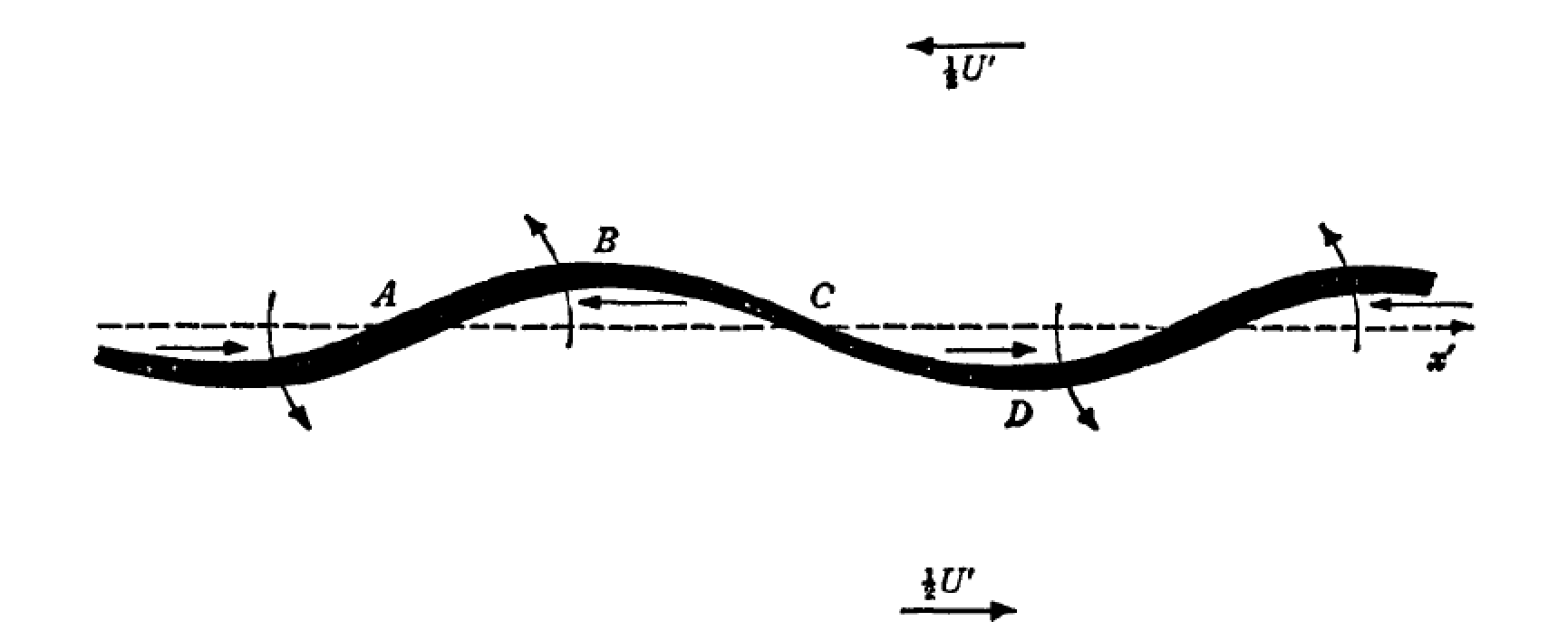}}
\caption{Instability of a sheet of vorticity submitted to an harmonic perturbation. The arrows show the velocities induced by the vorticity sheet on itself, leading to the reinforcement of the perturbation \cite{Batch1967}.}
\label{Batch}       
\end{figure*}

This is realized for example when two parallel streams with different velocity are superposed and create a turbulent mixing layer \cite{Thorpe1971}.
The instability mechanism has been described physically by \cite{Batch1967} by considering a sinusoidal perturbation of the vorticity sheet formed 
by the superimposed streams.
The sheet being a material element, a
perturbation $\propto e^{i (k_x x-\omega t)}$ with $k_x$ the horizontal wave number and $\omega = \omega_r + i \sigma$,
will disturb the vorticity sheet as in Fig.~\ref{Batch}. To predict the evolution of the perturbation, the self-advection produced by 
this vorticity distribution is analyzed. Expanding the vorticity sheet into
elementary elements and considering the velocities induced by each of these elements at point A (or C), 
we see that by symmetry the added velocities at point A vanish.
On the other hand, the velocities induced at the point B do not be compensate and will cause a horizontal 
displacement (to the left as shown on the figure) of the corresponding vorticity element. This process reinforces 
the local vorticity in A rather than in C. The asymmetric contribution from A and C on B will consequently displace the B element further up, leading
to the growth of the perturbation.
 
The rigorous mathematical treatment consists in applying continuity conditions at the interface and 
retaining solutions that vanish at large vertical distances. It shows that all disturbances are unstable with
a growth
rate $\sigma =  k_x \Delta U / 2 $ that depends on the horizontal wavelength of the perturbation
(Note that the dependence on $k_x \Delta U$ could have been anticipated from dimensional analysis).

\begin{figure}[h]
\centering
\resizebox{\hsize}{!}{\includegraphics{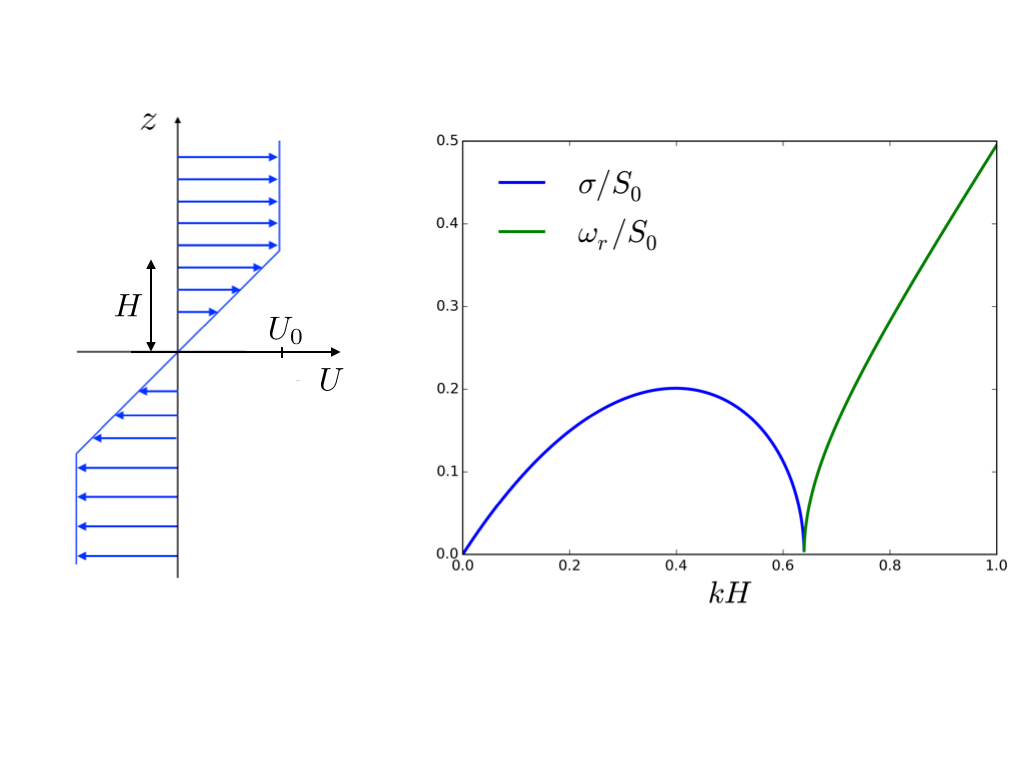}}
\caption{Piecewise linear shear (left). Growth rate as a function of the horizontal wave number of the perturbation (blue), oscillation frequency of propagative waves in the stable regime (green).}
\label{Schema2}       
\end{figure}

The absence of conditions on the lengthscale of the perturbation is due to the velocity jump between the streams.
We now consider the case of a continuous piecewise linear velocity
profile
presented in Fig.~\ref{Schema2} (left panel) where the shear has a proper lengthscale $H$ that will constrain the wavelength of the unstable modes.
This configuration also allows us to present another physical interpretation of the shear instability.
Again using continuity conditions at the lower and upper fluid interfaces $z = \pm H$, the 
dispersion relation of the perturbations reads $\omega^2 = \frac{S_0^2}{4}\left[ (1-2k_x H)^2 - e^{-4k_x H} \right]$,
where $S_0 = \Delta U/(2 H) = U_0/H $ is the shear rate.
Accordingly, the horizontal wavelength of the perturbation, $\lambda = 2 \pi/k_x$, must be sufficiently large $\lambda \ge 9.81 H $
to grow. The maximum growth rate is $\sigma_{max} \sim 0.2 S_0$ and it is reached for $k_{max} H \sim 0.40 $. In the stable $k_x H \ge 0.64$ regime,
the solutions of the dispersion relation are two waves propagating horizontally in opposite directions. In the
large $k_x H$ limit, their wave speeds take the simple form  $c_r = k_x \omega_r = \pm (U_0 - S_0/2 k)$ and their amplitudes are concentrated 
around the upper and lower interface, respectively. The solutions of the dispersion relation are displayed on Fig.~\ref{Schema2} (right panel). 

The long waves of the stable regime are known as vorticity waves (also called shear-Rayleigh or Rossby waves) that are in general present at vorticity jumps.
For example,
at the interface between
a layer of uniform velocity and a linearly sheared layer, $U(z) = \{ U_0 \; \mbox{for} \;  z>0 \; ;
\; S_0 z + U_0 \; \mbox{for}\;  z<0 \}$,
a vorticity wave propagates upstream at a phase velocity $c = U_0 -S_0 / (2k) $ and its amplitude
decreases as $e^{-k_x |z|}$ away from the vorticity jump.
Baines \& Mitsudera \cite{Baines1994} proposed an interpretation of the shear instability based on the interaction between the 
two vorticity waves produced at the upper and
lower vorticity jumps of the piecewise shear flow.
First, the requirement that the velocity field induced by one vorticity wave at the level of the other interface 
is not negligible
imposes a condition on the horizontal
wavenumber, i.e. $k_x H \lesssim 1$ since the wave amplitude decreases as $e^{-k_x |z|}$. Then, a phase-locking condition
between the two waves should ensure a sustained interaction between them. Imposing 
the same phase velocity for the two waves reads $U_0 -S_0 / (2k) = -U_0 +S_0 / (2k)$ leading to $k_x H = 1/2$.
These two physical conditions are reasonably close to the exact conditions for instability and maximum growth rate given above. 
See \cite{Baines1994}
for more details on how the waves reinforce each other.
Other instabilities like the baroclinic \cite{Vallis2006}, the Holmboe \cite{Baines1994} or the Taylor-Caufield \cite{Caulfield1995}
instabilities can be also interpreted as due to the resonant interaction between waves supported 
by the system.

For general profiles ${\bf U} = U(z) {\bf e_x}$ of inviscid parallel flows, the modal linear stability analysis provides necessary conditions for instability.
The Rayleigh criterion says that $U(z)$ must have an inflexion point inside the fluid domain,
while the stronger Fj{\o}rtoft criterion specifies, that, given a monotonic profile $U(z)$, 
$d^2U/dz^2 \left[U(z)-U(z_i)\right]$ must be negative somewhere in the domain, $z_i$ being the location of the inflexion point. 
In terms of the vorticity distribution, these criteria say there must be an extremum of vorticity in the domain (Rayleigh) and 
this extremum must be a maximum
of the (unsigned) vorticity (Fj{\o}rtoft) \cite{Drazin2004}. 

As known for a long time, however, 
experimental results on various shear flows show a transition to turbulence even when modal linear analysis finds stability. This holds for the
plane Couette
flow which is linearly stable for all Reynolds number and yet experimentally unstable 
for Reynolds numbers higher that $\sim 350$. The plane Poiseuille flows is linearly unstable
above a critical Reynolds number of $5772$ but experiments show transition to turbulence at a much lower Reynolds number $\sim 1000$.
To understand these discrepancies, a first step is to recognize that modal linear stability
does not guarantee 
that perturbations
monotonically decrease over time. 
In general, the evolution of perturbations can be put in the form 
$\frac{\partial}{\partial t}  \hat{u} = {\cal L}(\hat{u})$ where ${\cal L}$ is a linear
operator acting on the spatial part of the perturbation $\hat{u}$.
The modes are the eigenvectors of the linear operator and in the modal analysis the system is linearly stable
if the imaginary part of all the eigenvalues $\omega$ is negative.
 When ${\cal L}$ does not commute with its adjoint operator, 
the linear stability operator is said to be nonnormal and for shear flows, it is
generically the case. 
Then, even if modal analysis says the system is linearly stable, 
the kinetic energy of small initial perturbations can grow during a 
transitory phase before decreasing exponentially over larger time (in the linear approximation).
The transient growth is a consequence of the non-orthogonality of the set of eigenmodes of 
nonnormal operators \cite{Grossmann2000,Schmid2007,Schmid2012}. 
On the other hand, the linear stability operators of the centrifugal instability or of the Rayleigh-Bénard convection are normal and
in these cases, the critical parameters derived from the modal analysis correspond
to the experimental ones.

The maximum energy gain of the perturbations during the transient growth has been investigated for different shear flows
and it was found that is can be very large \cite{Trefethen1993}.
Thus, even if perturbations initially behaves linearly, the transient growth will induce
non-linear interactions which can lead to instability.
A well documented mechanism of non-modal energy growth in shear flows
is the lift-up process whereby counterrotating streamwise vortices generate growing streamwise velocity streaks.
It is part of a generic self-sustained mechanism for shear turbulence proposed in \cite{Hamilton1995, Waleffe1997}.
Accordingly, 
the streaks induced by the lift-up mechanism
are then unstable to three-dimensional instabilities and non-linearly regenerate the streamwise vortices.
The transition to turbulence in shear flows involving nonnormal amplification and non-linear interactions
is an active field of research in fluid dynamics \cite{Eckhardt2018}. In astrophysics this type of transition has been 
considered to investigate magnetorotational dynamos in accretion disks \cite{Rincon2007}.
Before the origin of these non-modal and non-linear instabilities were elucidated, J.P. Zahn \cite{Zahn1974} 
invoked the experimental results on shear flows as
an evidence that all shear flows are unstable above some critical Reynolds number $Re_c$,
of the order of $1000$.

\subsection{Stably stratified shear flows}
\label{sec-strat}

In an atmosphere stably stratified in the vertical direction, vertical motions necessarily come with a work of the buoyancy force. If the 
fluid elements are initially at their equilibrium level in the atmosphere, this work is negative and transforms
kinetic energy into potential energy. As a parallel shear flow instability induces
vertical motions, its development will be hindered by the stable stratification.

\begin{figure*}
\centering
\resizebox{\hsize}{!}{\includegraphics{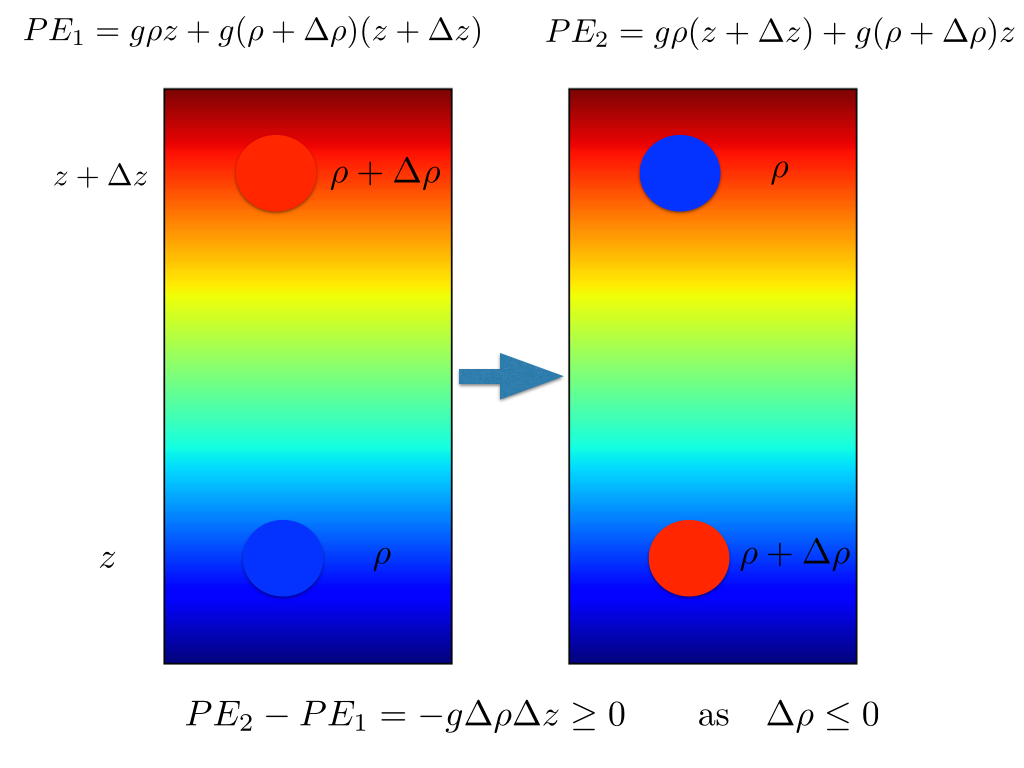}}
\caption{Potential energy required to exchange two fluid parcels in a stably stratified atmosphere}
\label{Strat0}
\end{figure*}

Following Chandrasekhar \cite{Chandra1961} (see also \cite{Miles1986}), the energy required to interchange two fluid elements can be compared to
the energy available in the shear in order to derive stability conditions.
Considering that the two fluid elements are initially at rest in the atmosphere
and are respectively located at $z$ and $z + \Delta z$, the difference in potential energy between the initial and the final states
is
$\rho_0 N^2 (\Delta z)^2$, where $N^2= - g/\rho_0 d\rho/dz$ is the Brunt-Väisälä frequency characterizing the background stratification.
As shown on Fig.~\ref{Strat0}, 
$Ep_2-Ep_1 =  g \rho(z+ \Delta z) +  g z (\rho+\Delta \rho) - \left[g \rho z + g (\rho+\Delta \rho)(z+ \Delta z)\right] = - g \Delta \rho \Delta z$
where $\Delta \rho <0$ if $\Delta z>0$ and $N^2 \sim - g/\rho_0 (\Delta \rho/\Delta z)$.
The energy that can be drawn from a shear flow is estimated
considering that the initial horizontal velocities of the fluid elements at $z$ and $z + \Delta z$
are respectively $U(z)$ and $U(z) + \Delta U$, and 
assuming a perturbation preserving horizontal momentum that brings both velocities to $U(z) + \Delta U/2$. The kinetic energy available
is thus given by
$Ek_2-Ek_1 = -\rho_0 (\Delta U)^2/4 $. The transformation will not be allowed
if the total, kinetic plus potential, energy of the system increases, that is if $Ek_2-Ek_1 + Ep_2-Ep_1 > 0$ which implies  $Ri = N^2/(\Delta U/\Delta z)^2 > 1/4$.
The stability of a parallel shear flow in a stratified atmosphere thus appears to be controlled by
a nondimensional number, called the Richardson number, that compares
the shear time scale $t_S = 1/(dU/dz)$ to the buoyancy time scale $t_B = 1/N$.

For an inviscid $\nu=0$ and diffusionless fluid $\kappa =0$, it is possible to obtain a rigorous derivation of this criterion.
Here we summarize the main steps of the calculation, the details can be found in \cite{Drazin2004, Rieutord2015}.
The basic state to be perturbed is again a general horizontal shear flow ${\bf U} = U(z) {\bf e_x}$ in a vertically stratified Boussinesq fluid with 
Brunt-Väisälä frequency $N(z)$ given by the thermal stratification $N^2 = \beta g dT/dz$ where $g$ is the gravity and $\beta$ the thermal expansion coefficient.
In the Boussinesq approximation \cite{Spiegel1960}, 
linear perturbations to this equilibrium are governed by
the following equations :

\beqa
\label{linear0}
\frac{\partial u}{\partial x} + \frac{\partial w}{\partial z} & = & 0\\
\frac{\partial u}{\partial t} + U \frac{\partial u}{\partial x} + w \frac{d U}{dz} & = & - \frac{1}{\rho_0} \frac{\partial p}{\partial x} + \nu \lapl u\\
\frac{\partial w}{\partial t} + U \frac{\partial w}{\partial x}  & = & - \frac{1}{\rho_0} \frac{\partial p}{\partial z} + \beta g \Theta + \nu \lapl w \\
\frac{\partial \Theta}{\partial t} + U \frac{\partial \Theta}{\partial x} & = & - w \frac{d T}{d z} + \kappa \lapl \Theta
\label{linear}
\eeqa
\noindent where $u$ and $w$ are the velocities in the streamwise and vertical directions
respectively, $p$ and $\Theta$ are the pressure
and temperature perturbations. Alternatively, the buoyancy perturbation field $b = \beta g \Theta$ can be used in place of $\Theta$, in which case
the Brunt-Väisälä frequency appears explicitly in the heat equation.

Neglecting viscosity and thermal diffusion, and looking for normal mode solutions $w(x,z) = \hat{w}(z) e^{i (k_x x-\omega t)}$, this set of equation can be reduced
to one equation, called the Taylor-Goldstein equation :
\beqa
\frac{d^2 \hat{\Psi}}{dz^2} + \left[ \frac{N^2}{(U - c)^2} - \frac{U''}{(U-c)} - k_x^2 \right] \hat{\Psi} = 0
\eeqa
\noindent where $\hat{\Psi}(z)$ is the complex amplitude of the streamfunction and $c = \omega/k_x$.
Introducing $\hat{\Phi}$, through $\hat{\Psi} = \left(\sqrt{U-c}\right) \hat{\Phi}$, multiplying the Taylor-Goldstein equation 
by the complex conjugate $\hat{\Phi}^{*}$, integrating across the fluid layer bounded by two rigid plates where the vertical velocities vanish, and taking
the imaginary part of the expression obtained, leads to :
\beqa
\sigma \left\{ \int_{\mathrm{bot}}^{\mathrm{top}}
\left(|\hat{\Phi}'|^2 + k^2 |\hat{\Phi}|^2\right) dz + \int_{\mathrm{bot}}^{\mathrm{top}} \left( N^2-\frac{U'^2}{4}\right) \left|\frac{\hat{\Phi}}{U-c}\right|^2  dz \right\} =0
\eeqa
As 
$\sigma \neq 0$ is a necessary condition for instability, the integrand must be zero which imposes that $N^2-\frac{U'^2}{4} < 0$ somewhere
in the fluid layer. In other words, a necessary condition for modal linear instability is :
\beqa
\mathrm{Ri} = \frac{N^2}{\left(\frac{dU}{dz}\right)^2} < \frac{1}{4} \;\;\;\; \mathrm{somewhere}\; \mathrm{in}\; \mathrm{the}\; \mathrm{fluid}
\eeqa
It is the so-called Miles-Howard criterion.

For the simple case of an hyperbolic-tangent parallel shear flow $U(z)= \Delta U \tanh(z/H)$ in a linearly stratified atmosphere $T(z)=T_0 + \Delta T z/H$, this
condition is found to be sufficient. Indeed, according to \cite{Drazin1958}, 
all perturbations with horizontal wavevector verifying $k_x H < 1$ and $Ri < (k_x H)^2 [1-(k_x H)^2]$ are
unstables. Thus, if the Richardson number is less than $1/4$, perturbations with $k_x H = \sqrt{2}/2$ will be unstable.

Before evaluating Richardson numbers in stars, we should mention that different types of instability  mechanism have been identified for 
sheared stratified atmosphere. In addition to the Kelvin-Helmholtz type which involves two interacting vorticity waves, two
other instabilities involving resonant interaction with gravity waves, the Holmboe and the Taylor-Caufield instabilities, have been identified.
The Holmboe instability
is relevant when the stable stratification varies on a length scale that is sufficiently smaller
than the lengthscale of the velocity gradient \cite{Smyth1988, Peltier2003}.
At first sight, this situation should not be generic 
in stellar radiative zones as angular momentum diffuses more slowly than heat. For example, in \cite{Witzke2015}, when considering an hyperbolic-tangent 
shear in a polytropic atmosphere, the Kelvin-Helmholtz instability was found to dominate over the Holmboe instability. 
It might nevertheless be relevant at the edge of evolved convective core where sharp compositional gradients are present.

Richardson numbers in stars can be estimated from the typical Brunt-Väisälä frequencies found in stellar evolution models
and from our knowledge of surface or internal rotation rates. 
For the solar tachocline, helioseismolgy provides us with both  the differential rotation across the layer $\Delta \Omega = 0.15 \Omega$
and an estimate of the layer thickness $H=4 \times 10^{-2} R_{\odot}$. Together with a typical value of the Brunt-Väisälä frequency
$N = 10^{-3} s^{-1}$, we get a very high Richardson number $Ri = \frac{N^2 H^2}{r^2 (\Delta \Omega)^2} = 2 \times 10^4$.
In the other cases where stellar seismology revealed differential rotation in radiative zones, in subgiants \cite{Deheuvels2014}
or in red giants \cite{Mosser2012}, the scaleheight of the rotation gradient is not known.
It is nevertheless reasonable to assume that the gradients take place over a distance of the order of the stellar radius. 
With this assumption, we can write $d U/d z \sim r \frac{d \Omega}{d r} = - \alpha \Omega$ with $\alpha= -\frac{d \ln \Omega}{d \ln r} \sim 1$.
Thus the Richardson number reads $Ri \sim \frac{N^2}{\Omega^2}$ and we can estimate its value for the Sun and for a typical early-type star.
For the Sun, using $\Omega = 2.76 \times 10^{-6} \mathrm{rad/s}$ and the previous value of $N$, we obtain $Ri = 1.3 \times 10^5$.
For a main-sequence intermediate-mass star with a period of rotation of $1$ day and a Brunt-Väisälä frequency $N = 10^{-3} s^{-1}$,
taken from stellar structure models of \cite{Talon2008}, we get $Ri = 2 \times 10^2$.
These estimates show that Richardson numbers in stellar radiative zones are typically very high. The Richardson
criterion for instability is thus very unlikely to be verified unless for some reason the rotation rate varies 
significantly over much smaller radial distance. 

\subsection{Stably stratified shear flows in highly diffusive atmosphere}
\label{lign99}

In the previous section, we considered a parallel shear flow in a stably stratified atmosphere and found
that, neglecting viscosity and thermal diffusion, stable stratification should hinder the shear instability in typical stellar
radiative zones.
However, we left aside many physical ingredients that may play a role in stars, including
rotation, sphericity, viscosity, thermal diffusion, density stratification, compressibility, magnetic field.
Among them, thermal diffusion 
can significantly alleviate the stabilizing effect of the stratification. Indeed, the buoyancy force $g \rho'/\rho$ is equal to $g \beta \Theta$ in the Boussinesq
approximation. By damping temperature deviations, thermal diffusion thus also
reduces the amplitude
of the buoyancy force and this will favor vertical motions.
At the same time, thermal diffusion is known to damp gravity waves. Thus it does not always favors vertical motions in stably stratified fluid.
The efficiency of these dynamical effects will depend on the motion lengthscale $\ell$ since the thermal diffusion time scale is $\ell^2/\kappa$ while the restoring
buoyancy force acts on a time scale $1/N$.
Interestingly, in the presence of a strong thermal diffusion,
it is not a trivial matter to decide whether a fluid element thrown vertically from its equilibrium position will end up above or below the maximum level of
its adiabatic oscillations. We can anticipate that the dynamical effect of thermal diffusion will depend on the time scale 

Thus, before we study how thermal diffusion affects the stability of stably stratified parallel shear flows, let us first clarify this point by considering
the simple case of small perturbations in an otherwise quiescent atmosphere in hydrostatic equilibrium.
The Brunt-Väisälä frequency is taken uniform so that modal perturbations $\propto e^{i(k_x x + k_z z - \omega t)}$ are also harmonic in the vertical direction,
where as before $\omega = \omega_r + i \sigma$.
The linear Boussinesq equations Eqs.~(\ref{linear0}-\ref{linear}), with $U$=0 and $\nu$=0, lead to
the dispersion relation : $\omega^2 + i \tau \omega - \omega_g^2 = 0$ where $\tau= \kappa k^2$ is
the thermal damping rate with $k^2 = k_x^2 + k_z^2$ and $\omega_g = \pm \frac{k_x}{k} N$ is the frequency
of adiabatic gravity waves.
The nature of the two solutions of this dispersion relation changes as thermal diffusivity is increased.
For low thermal diffusivity, they are damped gravity waves and, as long as the damping rate $\tau$ is significantly smaller than $\omega_g$, 
their frequencies $\omega_r \sim \pm \omega_g$ are practically not modified by
thermal diffusivity. Then, when thermal diffusivity exceeds some values, more precisely when $ 2 |\omega_g| \le \tau$,
the damping
is too strong to allow wave solutions. 
The perturbations are damped without propagation at a rate $\sigma^{\pm} = \frac{1}{2}(-\tau \pm \sqrt{\tau^2 - 4 \omega_g^2})$.
What is more interesting for our current purpose is the regime of still larger thermal diffusivity when the two
solutions $\sigma^{\pm}$ show very different behaviours. In the limit $ 2 |\omega_g| \ll \tau$,  
$\sigma^{-} = - \tau$ and $\sigma^{+} = -\omega_g^2/\tau$,
the first mode corresponds to a fast thermal damping that does not depend on the stratification while the second mode is also decreasing in amplitude
but with a damping rate that
decreases for larger thermal diffusivity and for smaller lengthscales. This latter mode characterizes the effect of the buoyancy force when it is strongly 
affected by the thermal diffusivity.
For isotropic perturbations, i.e. $k_x \sim k_z$, the damping time scale of this modified buoyancy reads
$t_{BM}= \kappa/(N^2 \ell^2)$. It can be expressed as a function of the
adiabatic buoyancy time
scale $t_B =1/N$  and the diffusion time scale $t_{\kappa}$, as $t_{BM}= t_B^2/t_{\kappa}$, that is,
the adiabatic buoyancy time increased by the factor $t_B/t_{\kappa}$ which is necessarily larger than one in the regime considered, $ 2 |\omega_g| \ll \tau$.

\begin{figure*}
\centering
\resizebox{\hsize}{!}{\includegraphics{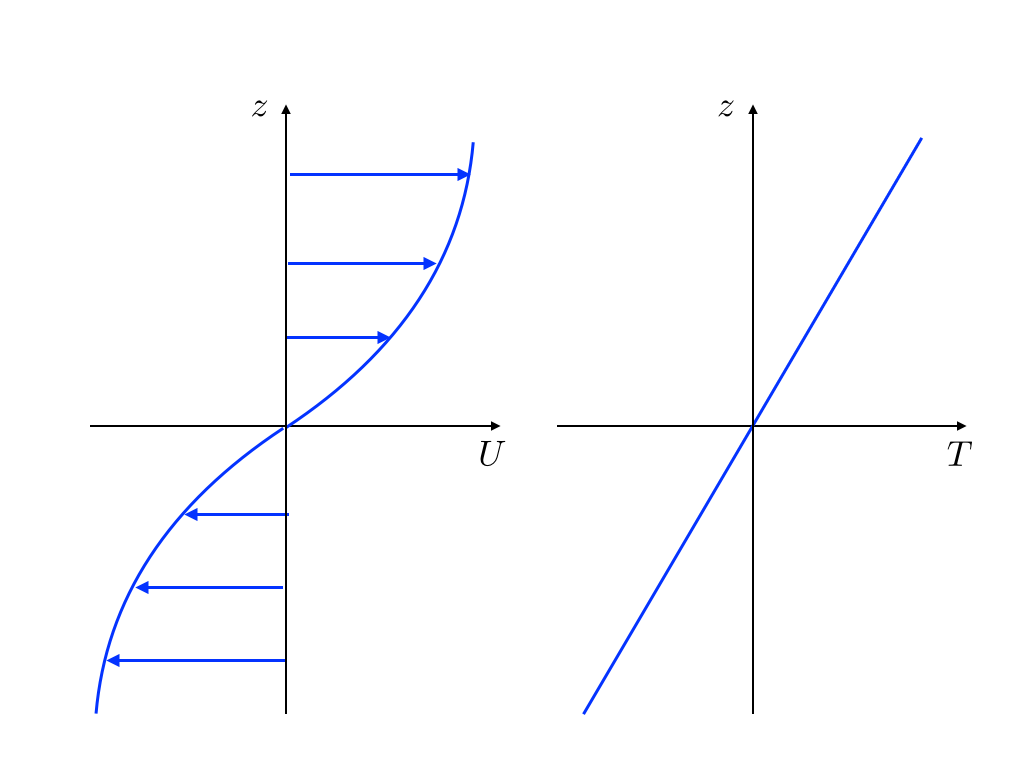}}
\caption{Inflectional parallel shear flow (left) in a linearly stably stratified atmosphere (right)}
\label{Schema5}
\end{figure*}

We now consider the effect of thermal diffusion on the modal linear stability of a stably stratified Kelvin-Helmholtz shear layer characterized
by an hyperbolic tangent profile $U(z)  = \Delta U \tanh (z/H)$ in a uniform Brunt-Väisälä frequency background.
For perturbations $ \propto \hat{w}(z) e^{i(k_x-\omega t)}$, the linear equations~(\ref{linear0}-\ref{linear}) reduce to a vertical 1D boundary value problem 
that is solved numerically.
This problem was studied by Lignières et al. \cite{Lign99} but also with minor variations by Dudis \cite{Dudis1974} and Jones \cite{Jones1977}. 
Witzke et al. \cite{Witzke2015} considered the effects of compressibility using a polytropic background.

\begin{figure*}
\centering
\resizebox{\hsize}{!}{\includegraphics{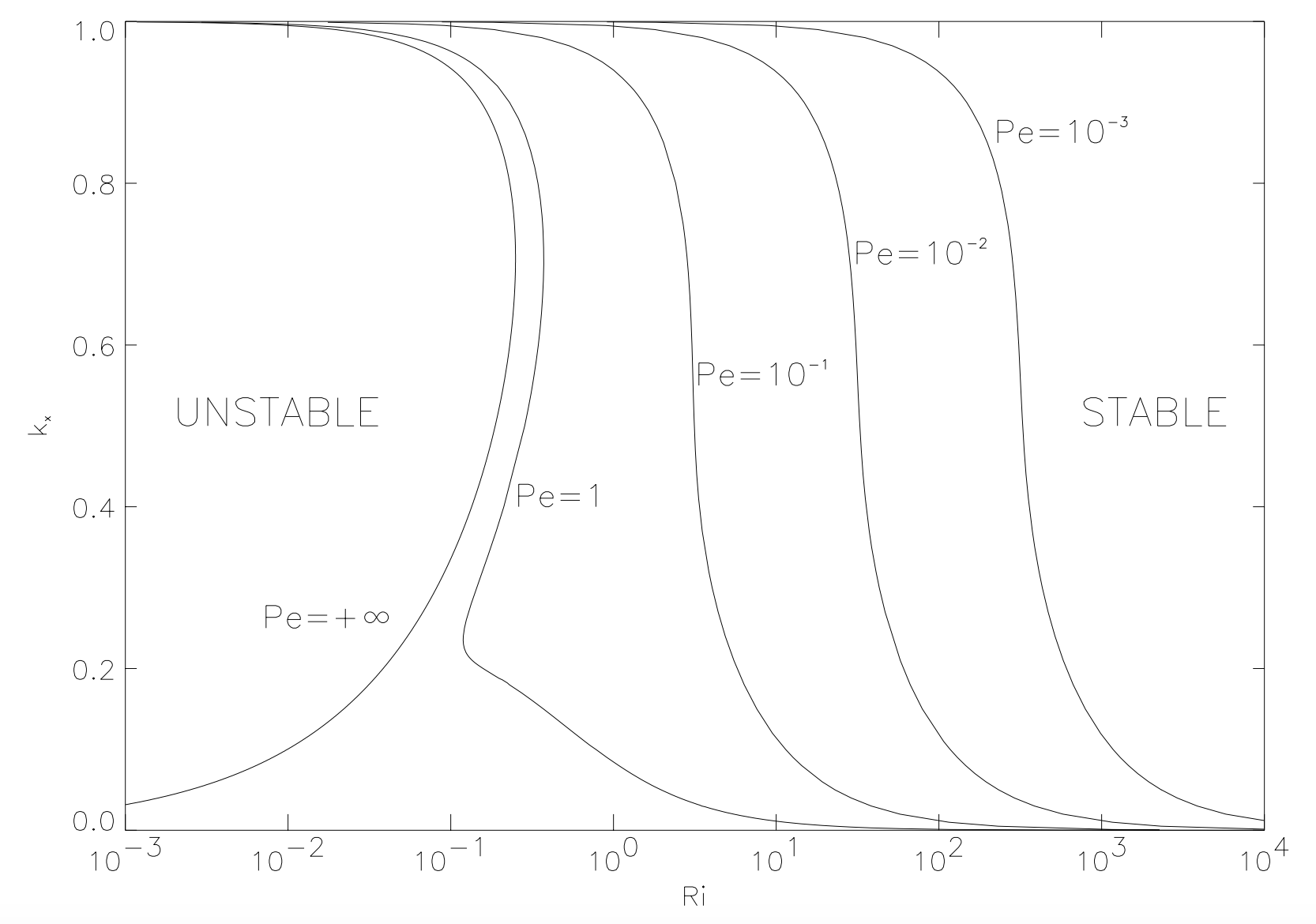}}
\caption{Effect of the thermal diffusivity on a Kelvin-Helmholtz instability in a linearly stratified medium. Neutral stability curves $k_x H = f(Ri)$, relating
the horizontal wavenumber of the perturbation $k_x$ scaled by the shear lengthscale $H$ to the Richardson number, delimit stable vs unstable domains 
for different values of the Peclet number.}
\label{omega_lign}
\end{figure*}

Figure~\ref{omega_lign} displays, in the inviscid case $\nu=0$, the neutral stability curves delimiting the stable and unstable domains as a function of 
the Richardson number $Ri = N^2 H^2/(\Delta U)^2$, the Peclet number $Pe= \Delta U H/\kappa$, and the horizontal wavenumber of the perturbation 
($k_x$ on the figure corresponds to $k_x H$ in the present notation). One observes that the instability occurs at large Richardson numbers if
the Peclet number is small enough.
This can be interpreted by comparing the three time scales involved, i.e. the buoyancy time scale $t_B = 1/N$, the  thermal diffusion time scale
$t_{\kappa}=H^2/\kappa$ and 
the shear time scale $t_S = H/(\Delta U)$.
According to the piecewise shear studied in Sect.~\ref{sec-shear}, the maximum growth rate of 
shear instability is $\sim 1/t_S$ and this 
occurs for a perturbation such that
$k_x H \sim 0.5$.

Thermal diffusion will 
play a role on the dynamics only when buoyancy has an effect on the dynamics. This last condition is verified
when 
$ t_B <t_S $ or $ Ri> 1 $.
Moreover, we have seen that thermal diffusivity significantly reduces the stabilizing effect of buoyancy when $t_{\kappa} <t_{B} $.
Thus, thermal diffusivity will affect the dynamics if $ t_{\kappa} \le t_B \le t_S$.

\begin{figure*}
\centering
\resizebox{\hsize}{!}{\includegraphics{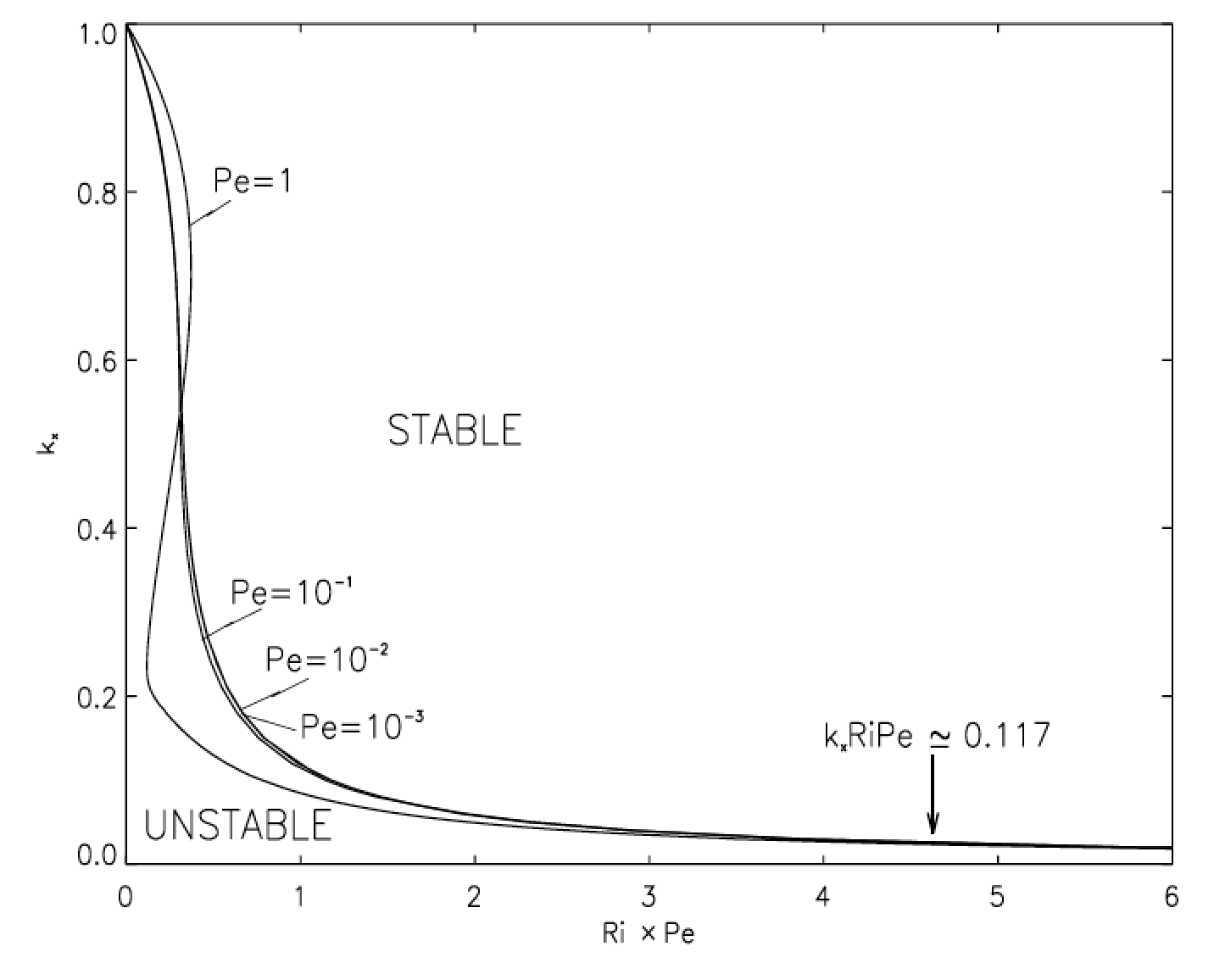}}
\caption{See previous figure. The stability curves are now displayed as function of the product $RiPe$.}
\label{omega_lign_bis}
\end{figure*}

It is possible to go further by using the concept of the modified buoyancy force introduced previously.
Assuming that in the regime $t_{\kappa} <t_{B}$, the combined effect of buoyancy and diffusion acts on a $t_{BM} $ time scale,
the stability threshold can be determined by comparing $t_{BM}$ and $t_S$.
Applying the condition $ t_S \leq t_{BM} $,
we obtain the criterion $ Ri Pe \leq 1$ for instability. This is essentially the correct result
as shown on Fig.~\ref{omega_lign_bis}, where the results presented in the previous figure are now plotted
in terms of the product $RiPe$.
Modes with $k_x H \sim 0.5$ are unstable for $RiPe \lesssim 1$ and their growth rates are of the order of $1/t_S$.

However, this figure also shows that disturbances with very large horizontal scales are unstable in the domain  $RiPe> 1$, the
condition on the horizontal wavenumber being $k_x H  <0.117 / (RiPe) $ \cite{Lign99}.
This low $k_x H$ tongue actually corresponds to a different type of modes that benefit from the fact that mostly horizontal motions
are practically  not affected by the  modified buoyancy. However, this set of modes must be considered
with caution for astrophysical applications. First, their growth rate is much smaller than the dynamical
one as $\sigma_{max} = 0.005 t_S /(RiPe)$. In addition, their horizontal lengthscale is so large that it may exceed the star radius.
Indeed, taking into account viscosity, 
\cite{Lign99} shows that $RiPe$ must be smaller than $1.67 \times 10^{-3} Re$ for instability and that the most unstable modes
have a ratio of the horizontal to the vertical lengthscale equal to $\sim Re/20$.
Applying this constraint to the solar tachocline values leads to a horizontal lengthscale of the unstable mode much larger that the solar radius,
showing that the unstable modes in the $RiPe > 1$ regime are not relevant for the tachocline.
We note also that the low $k_x H$ tongue is not always present as for example in  \cite{Dudis1974} where the temperature
background is an hyperbolic tangent profile instead of the linear one considered here.

I take this opportunity to make a side remark on the use of the term "secular" to describe instabilities in the stellar physics literature.
This terms in general refers to long time variations as compared to shorter periods present in the system (e.g. in celestial mechanics)
and for shear instabilities it is appropriate if the time scale of the instability is much larger than the dynamical time scale $t_S$.
This holds 
to describe the mostly horizontal modes of the $k_x H \ll 1$ tongue with growth rates $\sigma_{max} \ll 1/t_S$. The Goldreich–Schubert–Fricke instability \cite{Goldreich1967}
where
mostly horizontal modes strongly affected by thermal diffusion are also present is another
example where the term secular is appropriate.
But the growth rate of the $k_x H \sim 1$ modes being of the order of $1/t_S$, the present shear instability should not be called
secular, as it is usually done. 
In a stellar context, the difference between an adiabatic shear instability at $Ri<1/4$ and
and a diabatic shear instability at $Ri>1$ is that the vertical length scale of the latter has to be small enough to benefit from
the destabilization effect of thermal diffusivity, but their time scales are
of the same order. 

Modal linear stability shows that a Kelvin-Helmholtz shear layer is dynamically unstable whenever $Pe Ri \lesssim  1$. In an attempt to generalize these linear results
(at the time due to \cite{Townsend1958} and \cite{Dudis1974}) to the non-linear instability of general shear layers, 
Zahn \cite{Zahn1974} added the constraint that the Reynolds number should exceed some critical value $Re > Re_c$ (see Sect.~\ref{sec-shear}). 
As by definition $Pe=Re Pr$, these two conditions lead
to $Ri Pr < 1/Re_c$ as the condition for instability.
For stellar applications, Zahn proposed to take $Re_c \sim 1000$ from the transitional Reynolds observed in laboratory experiments (Couette,  Poiseuille flows),
that is $RiPr < 10^{-3}$ for instability.
Numerical simulations of shear-driven turbulent flows using different set-ups, homogeneous shear turbulence in Prat et al. \cite{Prat2016}, Couette flow 
in Garaud et al. \cite{Garaud2017},
both found that $Ri Pr$ must be lower than $0.007$ for the turbulence to be sustained. They confirm the existence of a critical $RiPr_c$ though somewhat larger than Zahn's estimate.
These simulations also study the turbulent transport induced by unstable shear flows. Their results will be described with more details in Sect.~\ref{shearpr}.

\section{Turbulent transport}
\label{turb}

As commented in the introduction, the turbulent transport of chemical elements and angular momentum in stars is modelled
through turbulent diffusivities associated with each identified instabilities. In the following we first recall what are the basis of the turbulent diffusion model
and most importantly the quite restrictive conditions for its validity Sect.~\ref{eddy}. These conditions can nevertheless be met and we present two examples
in Sect.~\ref{examples} that concerns 
the vertical transport of tracers in stably stratified turbulence at $Pr \sim 1$. How the associated eddy diffusivity 
relates
to the flow properties is discussed in Sect.~\ref{modelling}. As compared to the Earth atmosphere or the oceans, a specificity of the stellar fluid is
its very low Prandtl number meaning that
at microscopic scales heat diffusion is much more efficient than momentum diffusion. This is taken into account to study the vertical transport 
driven by shear turbulence in stellar conditions (Sect.~\ref{shearpr}) and to comment on the case where a mostly horizontal turbulence is forced on large horizontal scales
(Sect.~\ref{strongturb}).

In Tab. \ref{table1}, in addition to Prandtl numbers, some dynamical parameters of the stellar radiative zones and the Earth atmosphere and ocean are compared. 
Except in the convective boundary layer, motions in the atmosphere and in the oceans are typically strongly affected by the stable stratification.
The time scale of the 
observed large scale horizontal motions can be compared with the buoyancy time scale through
the horizontal Froude number $Fr_h = u_h/(N \ell_h)$, where $u_h$ and $l_h$ are the horizontal velocity and horizontal
length scale of the large scale motions. As displayed in Tab.\ref{table1}, these horizontal Froude numbers are very low
in the oceans and in the atmosphere
which means that the turbulence is strongly affected by the stable stratification.
We can not estimate Froude numbers from observations in stars.
On the other hand, we know from the ratio $N/\Omega$, also displayed in Tab. \ref{table1}, that 
in stars also the effect of stable stratification on vertical motions is dominant as compared to
that of the Coriolis force. 

\begin{table}
        \centering
        \caption{Relevant parameters of different stably stratified and rotating atmospheres.}
        \label{table1}
        \begin{tabular}{lccccccccccr} 
                \hline
                 & $N/\Omega$ & $l_h/u_h$ (days)  & $Fr = u_h /(l_h N)$ & $Pr$ \\
                \hline
                \, ~ \, Earth atmosphere \, ~ \, &  \, ~ \,  150  \, ~ \,& \, ~ \, 1 \, ~ \, & \, ~ \, $10^{-3}$ \, ~ \, &  \, ~ \,$0.7$ \, ~ \, \\
                \, ~ \, Earth ocean \, ~ \, & \, ~ \, 150  \, ~ \, &  \, ~ \, 10  \, ~ \, &  \, ~ \, $10^{-4}$  \, ~ \, &  \, ~ \, $\sim 10$  \, ~ \,\\
                \, ~ \, Sun \, ~ \, &   360 &   ?  &  ? &   \, ~ \,$10^{-6} - 10^{-5}$ \, ~ \, \\
                \, ~ \, Intermediate-mass star ($3 M_{\odot}$)\, ~ \, & \, ~ \, 14 \, ~ \, & ? & ? & \, ~ \, $10^{-7} - 10^{-6}$ \, ~ \,  \\
                \hline
        \end{tabular}
\end{table}

\subsection{The eddy diffusion hypothesis}
\label{eddy}

A basis for the eddy diffusion hypothesis is the work of G.I. Taylor \cite{Taylor1921}, who considered 
the dispersion of tracers 
in a turbulent flow under the assumption that the Lagrangian velocities
are statistically homogeneous. 
The main steps of his model are recalled below as this is useful to 
understand the limitations of the eddy diffusion concept.

It consists in relating the dispersion of fluid particles to the correlation of the velocity field.
Here we focus on $\langle [z(t)-z_0]^2 \rangle$, the mean square vertical displacement of fluid particles
from their intitial position $z_0=z(t=0)$. From $dz/dt = W$, the individual displacement is given by $z(t)-z_0 =\int_0^t W(t',z_0) dt'$ where
$W(t,z_0)$ is the Lagrangian vertical velocity. The mean square displacement then reads :
\beqa
\langle [z(t)-z_0]^2 \rangle & =& \langle W^2 \rangle \int_0^t \!\! \int_0^{t} \!\! R(t',t") dt' dt" 
\label{Tay1}
\eeqa
\noindent where 
\beqa
R(t',t'',z_0)  &=& \frac{\langle W(t',z_0) W(t",z_0) \rangle}{\langle W^2\rangle}
\eeqa
\noindent is the normalized autocorrelation of the Lagrangian vertical velocities.

Assuming statistical homogeneity of $W(t)$ allows us to limit the dependence of the autocorrelation to the time delay $\tau=t'-t"$.
The double integral can then be simplified (see details in \cite{Vallis2006}) to get :
\beqa
\langle [z(t)-z_0]^2 \rangle & =& 2 \langle W^2 \rangle \int_0^t \!\! (t- \tau) R(\tau) d\tau    
\label{Tay2}
\eeqa
A distinctive property of turbulence being its finite correlation time, one should be able to define
a Lagrangian correlation time $T_L$ as :
\beqa
T_L = \int_{0}^{+\infty} R(\tau) d\tau 
\eeqa
Depending on the time over which the dispersion is considered, two distinct regimes emerge from the expression~(\ref{Tay2}). A ballistic regime
at short time,  $t \ll T_L$, where the velocities are still well correlated and the mean dispersion is linear in time
$\langle [z(t)-z_0]^2 \rangle  = \langle W^2\rangle t^2$. Then,
over longer 
times $t \gg T_L$, the memory
is lost and the dispersion increases as if fluid parcels experienced a random walk :
\beqa
\langle [z(t)-z_0]^2 \rangle  = 2\langle W^2\rangle T_L t.
\label{Tay3}
\eeqa
By analogy, a diffusivity 
\beqa
D_t = \langle W^2 \rangle T_L 
\eeqa
is defined and is called turbulent (or eddy) diffusivity.

The $t \gg T_L$ condition imposes that a tracer field must vary on a sufficiently large length scale to experience eddy diffusion. 
Indeed, if $L_C$ denotes this length scale, the tracer distribution will evolve on a $L_C^2/D_t$ time scale and this has to be much larger than 
$T_L$. This is equivalent to $L_C \gg L_T$, where $L_T= W T_L$ is the  Lagrangian turbulent lengthscale associated with $T_L$.
On the other hand, if a tracer patch has a size smaller than $L_T$, its dispersion is described by considering pair dispersion \cite{Richardson1926,Sawford2001}.

Back to the condition on the turbulent velocity field, we note that Lagrangian statistical homogeneity is never strictly met in real flows.
In terms of Eulerian statistics, a flow that would be statistically stationary and homogeneous would satisfy Lagrangian statistical homogeneity.
But some level of Eulerian inhomogeneity is always present in real conditions for example near the boundaries of the system. 
Nevertheless, if 
the system has clear scale separation between the turbulent 
length scale $L_T$ and the lengthscale that characterizes the variation of the mean flow $L_S$, turbulent diffusion
should still be a good approximation to model the evolution of the tracer at the intermediate lengthscales.
These conditions really need to be checked for the particular flow considered.
For example,
the hypothesis of scale separation is far from being guaranteed. In
free shear flows like jets, wakes or mixing layer, the mean shear has a well-defined lengthscale which is also
the lengthscale of the largest eddies.

An eddy diffusion approximation can also be derived from the Eulerian equations using the mixing length theory.
The evolution of a conserved quantity $C$ in an incompressible turbulent flow with no mean velocity, $\langle {\bf u} \rangle =0$, is governed by :
\beqa
\derive{\langle C \rangle}{t}   &= - \nab \cdot \langle C'{\bf u'} \rangle + D_c \Delta \langle C \rangle 
\eeqa
\noindent where $C'$ are the fluctuation to the mean $C' = C - \langle C \rangle$ and $D_c$ is the molecular diffusivity. This equation simplifies to :
\beqa
\derive{\langle C(z) \rangle}{t}        &= - \derive{\langle C'w' \rangle}{z} + D_c \derived{\langle C(z) \rangle}{z}
\eeqa
\noindent when the mean distribution $\langle C \rangle $ only depends on $z$ and the turbulence is homogeneous in the horizontal directions.

The basic assumption of the mixing length theory is that a fluid parcel
carries its conserved quantity over a distance $\ell_{mix}$ before its mixes with the surroundings.
Consider a fluid parcel that moves from $z-\ell'$ to $z$ in the presence of a mean vertical gradient $d\langle C\rangle/dz$.
As $C$ is conserved, its deviation from the mean will be 
$C' = \langle C\rangle (z-\ell')- \langle C \rangle (z) = -\ell' d\langle C\rangle/dz + ..$ at $z$ before it mixes. 
If $\ell'$ is smaller than the scale over which the mean vertical gradient changes, the first order of the Taylor expansion is dominant.
The turbulent flux can then  be estimated by :
$\langle C'w' \rangle =  - \langle w' \ell' \rangle d\langle C\rangle/dz = - D_t d\langle C\rangle/dz$, where $D_t = \langle w' \ell' \rangle$ is a
turbulent diffusivity.

For generalized distributions $\langle C \rangle (x,y,z)$, 
$C' \sim -{\ell'}_z d\langle C\rangle/dz - {\ell'}_y d\langle C\rangle/dy-{\ell'}_x d\langle C\rangle/dx$ 
so that the three components of the turbulent flux $\langle C'{\bf u'} \rangle$ are written
$\langle u_i C' \rangle = - D_{ij} \partial_j  \langle C\rangle$  in terms of the components of a diffusivity tensor :
$D_{ij} = \langle u_i {\ell'}_j \rangle$.
We shall not consider here the implications of the tensor nature of the eddy diffusion as we investigate turbulent transport
in the radial (or vertical) direction in stars and we assume that turbulence homogenize the mean flow in the horizontal directions. However,
this aspect should be taken into account in 2D stellar models, where mean quantities depends on two spatial coordinates. 
We note that in this case the anti-symmetric part of the tensor acts as 
an additional meridional flow \cite{Vallis2006}.

In the mixing length theory, the two important hypothesis are the scale separation between the length of variation of the mean gradient of $\langle C \rangle$ and the mixing length, and 
the near conservation of the transported quantity, that is conservation except for the effect
of a small molecular diffusion. Irreversible mixing with the surrounding material is also important 
because the memory of $C'$ must be lost to ensure a significant mean correlation between the motion and the 
fluctuations of $C$.

Even if the diffusion hypothesis is correct, the value of the  eddy diffusivity still needs to be related to the flow properties.
The basic idea is to express it as $D_t = w_t \ell_{mix}$, 
where $\ell_{mix}$ is the mixing length and $w_t$ a r.m.s vertical velocity. But, to close the equations, the difficult part is 
to express 
both quantities as a function of the mean flow properties.
The mixing length theory has been first proposed by L. Prandtl \cite{Prandtl1925}
to describe the transport of horizontal momentum, a non conserved quantity, in shear flows.
The eddy viscosity $\nu_t = \alpha u_t \ell_{mix} $ is said to be proportional to the product of 
the mixing length $\ell_{mix}$ and a velocity scale $u_t$ estimated
by $\ell_{mix} |d\langle U\rangle /dz|$. In free shear flows like jets, wakes or mixing layers, the characteristic length scale of the mean shear flows 
is a natural choice for
$\ell_{mix}$. It turns out that this mixing length model provides reasonably accurate profiles of the mean velocity for calibrated value 
of the non-dimensional parameter $\alpha$, which varies 
between $0.07$ and $0.18$ for the three mentioned free shear flows \cite{Tennekes}.
Considering that both the assumptions of materially conserved quantity and scale separation are not verified in these cases,
the success of this mixing length model is somewhat surprising.
Tennekes \& Lumley \cite{Tennekes} attribute it to dimensional analysis as these flows are 
characterized by one length scale $L_S$
and one time-scale $1/(d\langle U\rangle/dz)$. 
In the general case however, and in particular when additional physical effects 
come into play, using dimensional analysis is not sufficient to prescribe how the eddy diffusivity
depends on the mean flow properties. This remark holds in particular when stable stratification adds 
a new time scale $1/N$ 
in a turbulent shear flow.

To summarize this part, previous works indicate that the eddy diffusion hypothesis can be a valid approximation to describe 
turbulent transport under certain circumstances \cite{Vallis2006}. Indeed, for a 
materially conserved quantity whose initial distribution $C_0(z)$ varies over length scales larger
than the turbulent lengthscale,
a turbulent flow with homogeneous Lagrangian statistics and
finite correlation time $T_L$, will act as a diffusion process on its mean distribution $\langle C \rangle (z,t)$.
Whether these conditions are met depends on the particular flow considered. Even so, there is no obvious choice 
in general to express the eddy diffusivity in terms of
the mean flow property. 

\subsection{Vertical eddy diffusion in stably stratified turbulence at $Pr \sim 1$}

In this section, we consider the problem of vertical turbulent transport in a stably stratified atmosphere. The turbulence
is feeded by some forcing mechanism, the stable stratification
is specified by the Brunt-Väisälä profile $N(z)$, and the fluid properties by a molecular viscosity and a thermal diffusivity.
Except for an example of turbulent transport in the ocean, we focus on results of numerical simulations. 
Indeed in numerical simulations, the forcing
mechanism can be chosen to generate statistically stationary and homogeneous turbulent flows which is a major advantage
to study turbulent transport. 
As implied by Eulerian homogeneity in space and time, the Lagrangian statistics is homogeneous and
the transport of a tracer if it occurs on time scale larger than the Lagrangian time scale $T_L$ should be described by an eddy diffusion.

In the following we illustrate with two examples 
that eddy diffusion can indeed be a good model for passive scalar vertical transport
in stably stratified turbulence (Sect.~\ref{examples}). Then, we comment on the physical mechanism behind 
this transport 
in terrestial conditions where $Pr\sim1$ (Sect.~\ref{modelling}). 

\subsubsection{Examples of vertical eddy diffusion}
\label{examples}

\begin{figure*}
\centering
\resizebox{\hsize}{!}{\includegraphics{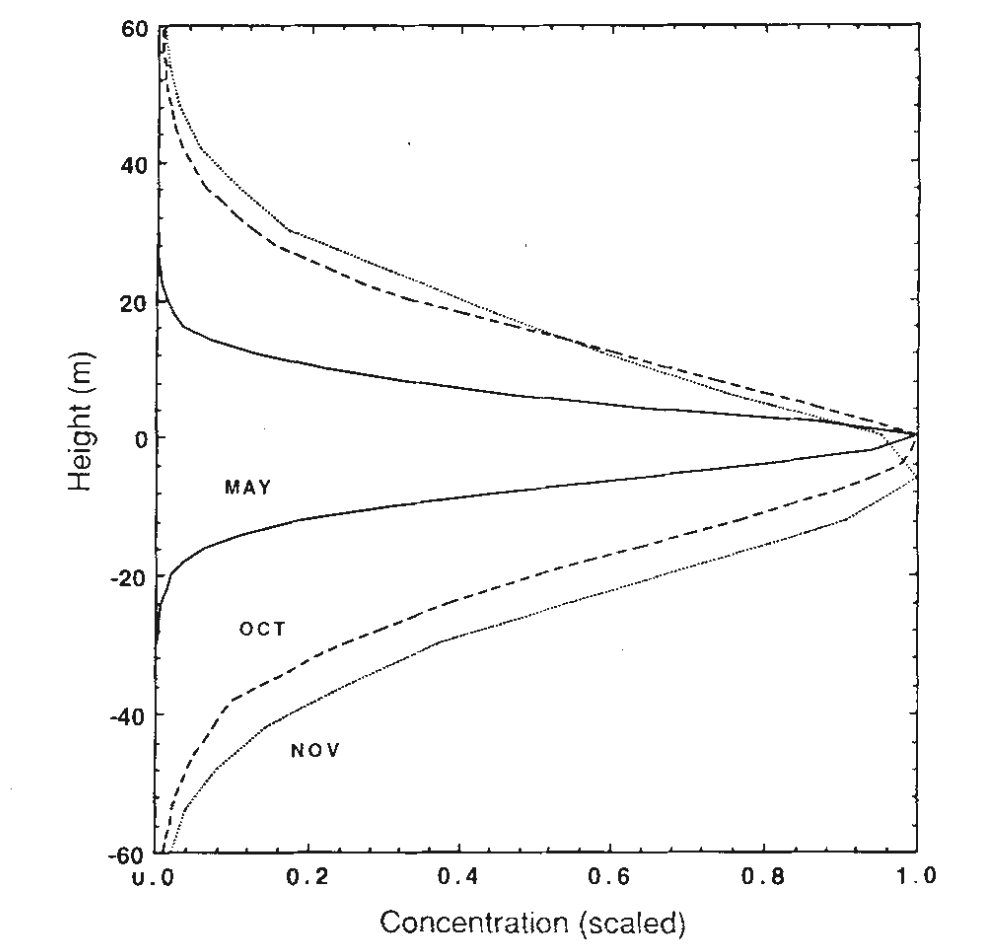}}
\caption{Vertical profiles of the concentration (scaled) of a tracer released at a depth of about $300 \mathrm{m}$ in the ocean and followed
during several months \cite{Ledwell1993}}
\label{Ledwell}       
\end{figure*}

While estimates of the vertical (or diapycnal to take into account the fact that isodensity surfaces are not strictly horizontal)
eddy diffusion in the oceans often rely on indirect measurements \cite{Gregg2018},
the release and following of tracers allow direct in-situ transport measurements and constitute a nice
illustration for our purpose.
In an experiment reported in \cite{Ledwell1993},
inert chemicals 
($139 \mathrm{kg}$ of dissolved SF6) were released in the ocean at a depth of about $300 \mathrm{m}$
and followed
over seven months.
The initial patch is dispersed mostly horizontally (or parallel to the density surfaces) but a slow vertical
dispersion also takes place as shown in Fig.~\ref{Ledwell} where the tracer vertical distribution is displayed at different epochs.
The increase of the width of a Gaussian-like profile 
resembles what is expected for a
diffusion process. 
The width is measured by the second moment of the concentration profile $M_2$
and its (almost linear) rate of growth provides a value for the effective vertical diffusivity using $2 D_{\rm eff} = (1/2) dM_2/dt$.

A value of $0.11$ \cms is found in this experiment. Much higher eddy diffusivity, up to $10$ \cms, have also been
measured by the same method over rough topography
in the abyssal ocean \cite{Ledwell2000}.

We now turn to numerical simulations of homogeneous stably stratified turbulence
where a random forcing term injects energy into large scale vortical motions (the forcing is concentrated around an
horizontal length scale equal to $L_f = L_h/4$, where $L_h$ is the horizontal size of the numerical domain).
The energy is transferred to small scale through a turbulent cascade where it is dissipated and a statistically stationary state is reached.
A uniform Brunt-Väisälä frequency and periodic boundary conditions in the three spatial directions
insure that the turbulence is statistically homogeneous \cite{Lindborg2009}.

\begin{figure*}
\centering
\centering
\resizebox{\hsize}{!}{\includegraphics{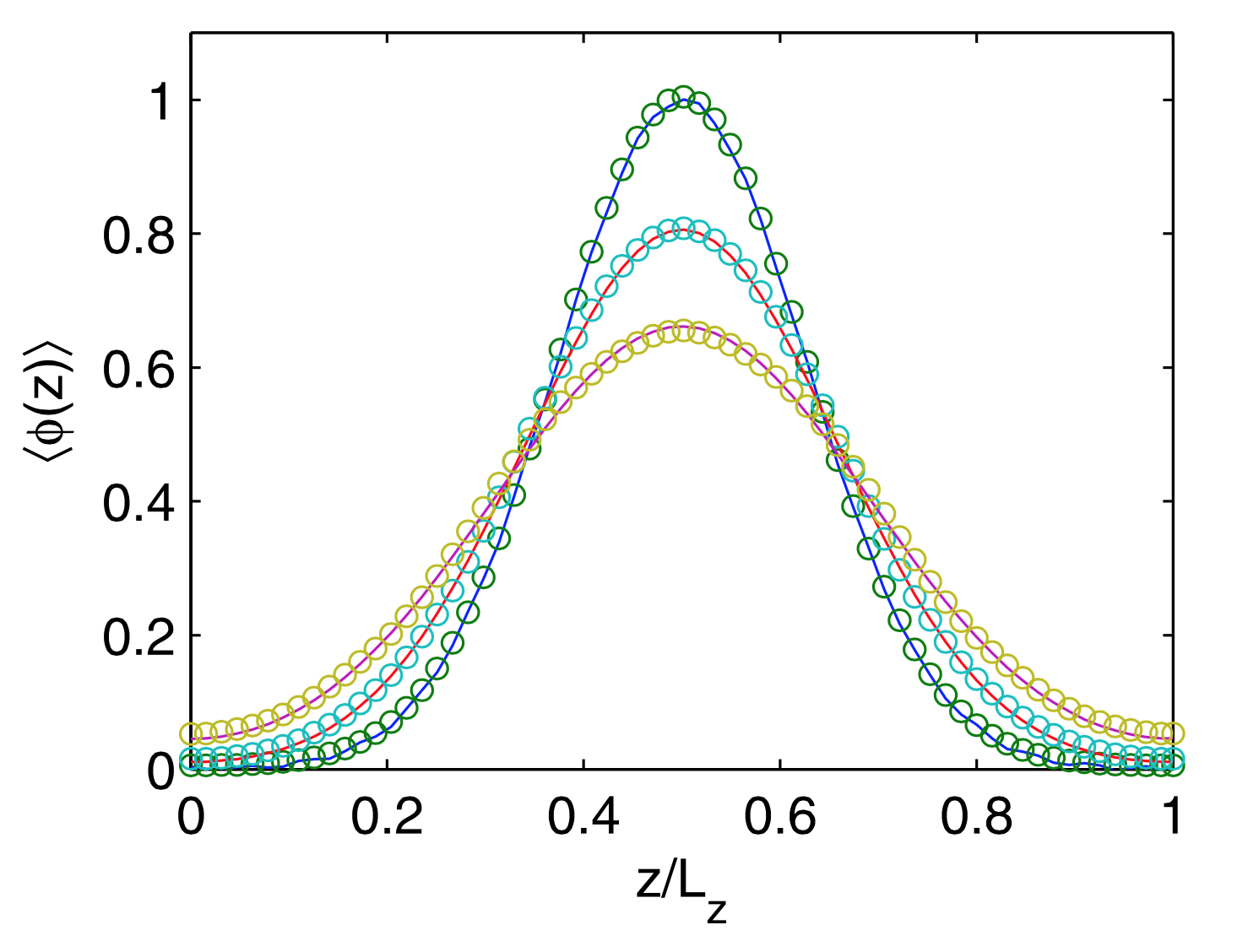}}
\caption{Vertical profiles of the mean concentration field of a tracer released in the midplane of a numerical simulation of randomly forced stationary
and homogeneous stably stratified turbulence. The initial vertical profile is Gaussian and the mean concentration appears to be diffused vertically. The solid curves
are analytical solutions of the 1D diffusion equation with $D_t = \epsilon_P/N^2$ \cite{Lindborg2009}.}
\label{Fedina}       
\end{figure*}

The advection-diffusion equation of a passive scalar field initially concentrated 
in a layer located at the mid-plane of the numerical domain
is then solved numerically. The initial field is homogeneous horizontally $C_0(z)$ and its vertical distribution
is a Gaussian
profile of width $h_0$. Figure ~\ref{Fedina} displays the initial profile together with the horizontal average of the scalar
field at two later times. The first observation is that again the vertical transport of the tracer 
apparently behaves like a diffusion as the width of the Gaussian increases in time.
Indeed, solving a 1D vertical diffusion equation with the initial condition $C_0(z)$ and using the diffusion coefficient $D_t= \epsilon_P/N^2$,
where $\epsilon_P = \kappa \langle \nab b \cdot \nab b \rangle$ is the mean dissipation rate of potential energy, 
closely models the evolution of the mean scalar
field in the numerical simulation (as shown by the superposition of the model and numerical results on Fig.~\ref{Fedina}).
We recall that $b = \beta g \Theta$ is the buoyancy fluctuation field.
Such a good agreement requires that the initial layer width $h_0$ is much larger than the vertical
length scale of the turbulence defined by $\ell_v= \sqrt{E_P}/N$, where $E_P = 1/2 \langle b^2 \rangle$ is the 
mean turbulent potential energy. This ratio is equal to $18.5$ for the simulations
shown on figure~\ref{Fedina}. When $h_0/\ell_v = 4.8$, there is a clear disagreement 
between the model and 
the simulation, which can not be improved significantly by tuning the eddy diffusivity (see details in \cite{Lindborg2009}). 

This example shows that a gradient diffusion model for the vertical turbulent transport can be excellent in stably stratified
turbulent flows. Stable stratification helps as it strongly constrains the vertical length of the turbulence 
and thus opens
a range of scales that are at the same time larger than the vertical turbulent lengthscale $\ell_v$ and 
smaller than the vertical lengthscale of the mean flow.
As mentioned above, such a scale separation is not often achieved in turbulent flows
like unstratified shear layer or thermal convection. 
In these simulations instead, the ratio between $\ell_v$ and the vertical size $L_z$ of 
the numerical domain can be controlled by tuning $L_z$ and the rate of energy injection by the random forcing.

This example also provides
an expression for the eddy diffusivity that contains no free parameter.
Its physical ground are exposed in the next section.

\subsubsection{Modelling the vertical eddy diffusion}
\label{modelling}

An important specificity of the vertical turbulent transport in stable atmosphere is to require some irreversible mixing to take place.
The vertical displacement of fluid parcels from their equilibrium position indeed produces buoyancy fluctuations that increase the mean potential energy $E_P$.
Therefore, if the process is adiabatic, a monotonically increasing vertical dispersion $\langle [z(t)-z_0]^2 \rangle$ comes with an ever increasing  
mean potential energy.
This would require a corresponding increase of the turbulent kinetic energy that is not possible if the turbulence is stationary.

A finite vertical dispersion of particle is indeed observed 
at intermediate times in stationary turbulence \cite{van2008}.
This might seem at odds with Taylor diffusion model but 
numerical simulations of slowly decaying homogeneous stably stratified turbulence \cite{Kaneda2000} show 
that the Lagrangian autocorrelation of the vertical velocity oscillates as a function of the time lag $\tau$ which leads to a vanishing Lagrangian correlation
time $T_L$ \cite{Kaneda2000}. Note that these strongly stratified flows have been modeled as a statistical superposition of gravity waves.

Thus, a monotonic vertical dispersion in a stationary turbulence must involve interchanges of density between fluid elements and in a turbulent flow this will occur 
through stirring and diffusive irreversible mixing at small length scales. 
Building on \cite{Pearson1983},
Lindborg \& Brethouwer \cite{Lindborg2008} proposed a model for the vertical dispersion in stationary turbulence that include these effects :
\beqa
\langle [z(t)-z_0]^2 \rangle
& = & \frac{\langle [b(t)-b_0]^2 \rangle}{N^2} + 2 \frac{\epsilon_P}{N^2} t
\label{mod}
\eeqa
\noindent where the first term increases with time up to a finite limit $4 E_P/N^2$, that corresponds to the maximum vertical dispersion in adiabatic conditions.
The second term dominates the first one for times larger than an eddy turnover time and corresponds to a vertical diffusion with eddy diffusion $D_t = \epsilon_P/N^2$.
According to the authors, the main assumption in their derivation is 
that the interaction time scale for exchange of density between fluid elements is
equal to the Kolmogorov time scale $(\nu/\epsilon)^{1/2}$, where $\epsilon$ is the mean dissipation rate of kinetic energy.

\begin{figure*}
\centering
\resizebox{\hsize}{!}{\includegraphics{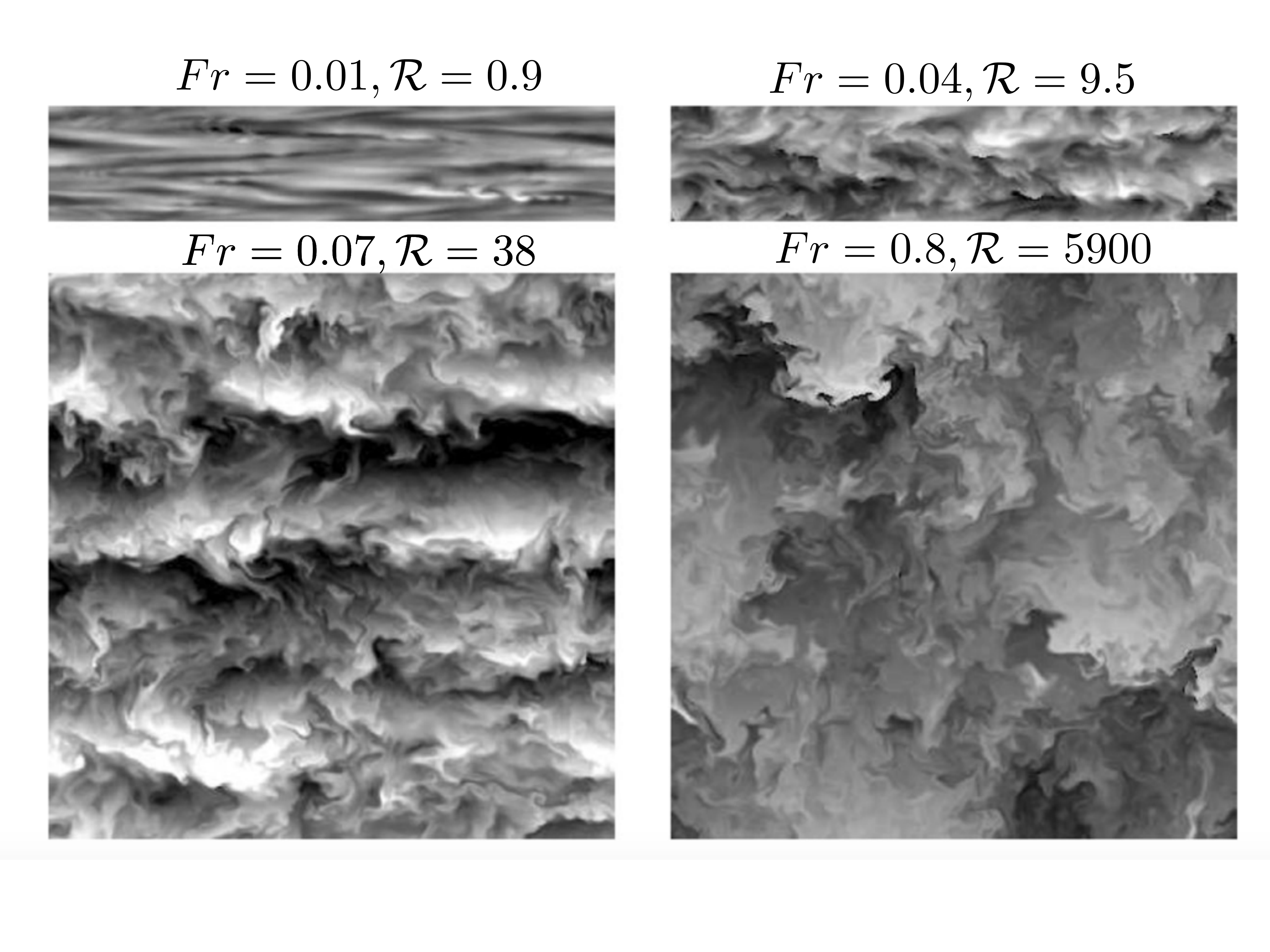}}
\caption{Snapshots of the buoyancy fluctuations in a vertical plane, for different simulations of randomly forced stably stratified turbulence \cite{Bretou2009}.}
\label{Bretou1}
\end{figure*}

This model has been tested in numerical simulations of homogeneous stably stratified turbulence sustained by large scale random forcing, like the ones
mentioned in the previous section \cite{Bretou2009}. The two main parameters of these simulations are the horizontal Froude number $Fr_h = u_h/(N l_h)$, that controls the buoyancy effects on the large horizontal scales, and the buoyancy 
Reynolds number ${\cal R} = \epsilon /(N^2 \nu)$, that controls the scale separation between the scales not affected by the stratification and the Kolmogorov scale
$(\nu^3/\epsilon)^{1/4}$.  
Snapshots of the buoyancy fluctuations in a vertical plane are shown in Fig.~\ref{Bretou1} for different values of these two parameters.
They show that the flow is more anisotropic as the Froude number decreases. At at low $Fr_h$, strong vertical shears develop between the horizontal layers 
and, if the Reynolds number is sufficient, they break down into 3D turbulence as seen in the ${\cal R} = 9.5$ and ${\cal R} = 38$ simulations.
In this case, the vertical length scale adapts to $\ell_v \sim u_h/N$, which explains the difference in the layer thickness between the $Fr_h=0.04$ and $Fr_h=0.07$ simulations. We shall come back to this point in Sect.~\ref{strongturb}.

The vertical dispersion computed in these different flows shows a good agreement with the model of Eq.~(\ref{mod}), especially for the vertical diffusion term and when
the buoyancy Reynolds number is large enough (see details in  
\cite{Bretou2009}). 

Actually this expression of the vertical eddy diffusivity is known from Osborn \& Cox \cite{Osborn1972,Osborn1980} who
derived it by assuming a balance between production and dissipation in the equation of conservation of the buoyancy fluctuations.
In the context of the Kolmogorov energy cascade, $\epsilon_P$ is independent of the molecular diffusion $\kappa$ and is instead related
to the dynamics. This led \cite{Osborn1980} to express the eddy diffusivity as a function of $\epsilon$ :
\beqa
D_t = \Gamma \frac{\epsilon}{N^2}
\eeqa
\noindent where $\Gamma = \epsilon_P/\epsilon$ is referred to as the "mixing efficiency".
The value of $\Gamma$ was assumed to be $0.2$ by \cite{Osborn1980} although it has since been shown to vary with the type of flow considered 
and with the Froude number \cite{Peltier2003,Maffioli2016,Qi2017}. Nevertheless, according to \cite{Maffioli2016}, 
it approaches a constant value $0.33$ at low Froude numbers.

\subsection{Vertical eddy diffusion due to radial shear in radiative zone}
\label{shearpr}

Here we consider the special case of shear driven turbulence in a stably stratified atmosphere with high thermal diffusivity.
We have seen in Sect.~\ref{lign99} that, in stars, for radial differential rotation with shear rate of the order of $\Omega$, stable stratification
inhibits shear instability unless thermal diffusion comes into play. 
Linear stability then showed that a strongly stratified shear layer ($Ri\gg 1$) can still be 
unstable on a dynamical time scale if the thermal diffusion time scale $t_{\kappa} = \ell_v^2/\kappa$ is smaller than the buoyancy time scale $t_B=1/N$.
This condition puts a strong constraint on the vertical length scale of the motions involved namely $\ell_v< \ell_c= \sqrt{\kappa/N}$.
A typical value for $\ell_c$ in the solar radiative zone is $1 \mathrm {km}$, which means we are considering very small length scales as compared
to the star radius. By the way, this tends to justify the use of the plane parallel geometry and of the Boussinesq approximation.

Zahn \cite{Zahn1992} proposed that the vertical turbulent transport driven by a vertical shear can be modeled by a turbulent diffusion of the form :
\beqa
D_t \sim (Ri_c/3) \kappa Ri^{-1}
\label{Za1}
\eeqa
\noindent with $Ri_c=1/4$.
The starting point of Zahn's model was the linear stability analysis of Dudis \cite{Dudis1974} stating, as above, that 
instability requires $RiPe < (RiPe)_c$.  He added that in a turbulent
regime it is  
the turbulent Peclet number $Pe_{\ell}=u \ell/\kappa$
that controls whether an eddy of size $\ell$ and velocity $u$ behaves adiabatically or not in the stably
stratified atmosphere. Then, the largest eddies authorized by the stability constraint are such that $Ri Pe_{\ell} \sim (RiPe)_c$.
The prescription Eq.~(\ref{Za1}) follows from the usual relation between the turbulent diffusivity and the eddy scale and velocity $D_t \sim u \ell/3$
(and using $Ri_c$ instead of $(RiPe)_c$).

In the following
we review numerical simulations designed to test Zahn's prescription. We can distinguish 
the scaling law $D_t \propto \kappa Ri^{-1}$ that contains the main physical assumptions of the model from the scale factor, 
$Ri_c/3$, that is not supposed to be better than an order of magnitude estimate.
In principle, numerical simulations or laboratory experiments of turbulent flows
are best suited to constrain such a factor.

However, the usual stellar conditions being $Ri \gg 1$, the consequence of Zahn's prescription is that turbulent flows with very
low Peclet numbers must be simulated. This in turn implies that the Prandtl numbers must be very low, which is indeed a
specificity of the stellar fluid.
Very low Prandtl numbers put strong constraints on laboratory experiments, because of the lack of low Prandtl fluids, but also
on numerical simulations.
Typically, the smallest time scale that needs to be resolved in numerical simulations of turbulent flows is the viscous dissipation at
lengthscales close to the spatial resolution. The numerical time step must then be
smaller than $\sim (\Delta x)^2/\nu$, with $\Delta x$ the spatial resolution.
But at low Prandtl number, the thermal diffusion time scale at this lengthscale is much smaller.
This 
means the numerical time step must be decreased by a factor $1/Pr$, which is huge for typical stellar Prandtl numbers.

\subsubsection{The small-Peclet-number approximation}
\label{smallpe}

This difficulty can be avoided by considering the Boussinesq equations in the limit of small Peclet numbers. The Boussinesq equations
are written around an hydrostatic solution, for which the thermal background is a solution of the heat equation. In a plan parallel
geometry, these equations read :
  \beqa
                        \grad\cdot{\bf v}                                       &=&0                                                                     \\
       \derive{\bf v}{t}+({\bf v}\cdot\grad){\bf v}            &=&-{\grad} p+Ri\Theta{\bf e}_z+\frac{1}{Re}{\bf \Delta}{\bf v} \\ 
          \derive{\Theta}{t}+{\bf v}\cdot{\grad}\Theta+w          &=&\frac{1}{Pe}\Delta\Theta	
\label{Bou}
  \eeqa
\noindent where the dimensionless numbers $Ri = (N^2 L^2/U^2)$, $Pe = (UL)/\kappa$, $Re=(UL)/\nu$ are expressed in terms of units of length $L$, velocity $U$ and 
the Brunt-Väisälä frequency $N^2 = \beta g \Delta T/L$ associated with
the temperature difference $\Delta T$ between the upper and lower horizontal limits of the layer.
All the fields are expanded in ascending powers of the Peclet number ($w = w_0 + Pe w_1 + ..., \; \Theta = \Theta_0 + Pe \Theta_1 + ...$).
At zero order, the solution of the heat equation is $\Theta_0=0$ which means that the thermal background remains unchanged. At the next order,
the Lagrangian time derivative of the temperature is negligible except for the term of vertical advection against the thermal background that
cannot disappear. 
Replacing the fluctuations of temperature $\Theta$  by a new variable $\Psi = \Theta/Pe$ then leads to the small-Peclet-number equations \cite{LignPe} :
   \beqa
\label{pet}
\grad\cdot{\bf v}                                       &=&0                                                                     \\
     \derive{\bf v}{t}+({\bf v}\cdot\grad){\bf v}            &=&-{\grad} p+RiPe\Psi{\bf e}_z+\frac{1}{Re}{\bf \Delta}{\bf v}\\ 		
	     w                                                    &=&\Delta\Psi
\label{petit}
  \eeqa
Without time derivative in the heat equation, there is no more constraint on the numerical time step associated with thermal diffusion.
But one should keep in mind that formal asymptotic expansions can be singular, a well-known example being the singular behaviours of the viscous dissipation
in the large  Reynolds number limit. 
This question has been investigated by mathematicians who confirmed the validity of the small-Peclet-number approximation \cite{Donatelli2016}.
Physically, the linear evolution of harmonic perturbations in a quiescent atmosphere that we considered in Sect.~\ref{lign99} provides some clues. We have seen that for large enough thermal diffusivity, 
pure thermal diffusion modes separate
from the dynamics. The small Peclet number expansion is a way to filter them out. 
This is reminiscent of the Mach number expansion that enables to filter out acoustic waves, leading
to the anelastic approximation \citep{Gough1969}.

The next interrogation is how small the Peclet number has to be for the approximation to be relevant.
For the Kelvin-Helmholtz instability described above, it was shown that the small-Peclet-number approximation is already 
very close to the Boussinesq calculations at $Pe=0.1$ \cite{Lign99}. Similar results have been obtained in the non-linear regime by \cite{Prat2013,Garaud2016}.

The small-Peclet-number approximation also has the advantage of simplifying the physical interpretation 
of the effects of the stable stratification and of the thermal diffusion. 
Both effects indeed combine in a single process, a buoyancy modified by thermal diffusion, with a time scale $t_{BM} = t_B^2/t_{\kappa}$.
In Eqs.~(\ref{pet}-\ref{petit}), this translates into the combinaison of two independent
non-dimensional numbers $Ri$ and $Pe$, into a single one $RiPe$ that controls the relative importance of the modified buoyancy with respect to the dynamics. 

Moreover, the energy conservation equation shows that the work of the buoyancy force always extract kinetic energy \cite{LignPe} (as already observed above with the
damped linear modes in an otherwise  quiescent atmosphere).
Thus, the concept of available potential energy, that is the energy stored
into buoyancy fluctuations that can be transformed back to kinetic energy, is no longer relevant in this limit.  

This property incidently clarifies a physical interrogation related to the fact that Zahn's eddy diffusion is proportional to $\kappa$.
On the one hand, a higher thermal diffusion indeed favors the vertical transport by reducing the temperature deviations thus also the amplitude of the buoyancy force.
But, the opposite and concomitant effect, namely that the kinetic energy
extracted by the buoyancy will be more efficiently dissipated by thermal diffusion, does not seem to be taken into account in Zahn's criterion.
This issue is solved in the small-Peclet-number regime because, as we just mentioned, all 
the kinetic energy extracted by the buoyancy force
is immediately and irreversibly lost thus, obsviously, increasing the thermal diffusion cannot increase the dissipation of kinetic energy.
In this regime, increasing thermal diffusion just favors the dynamics which is in agreement with Zahn's eddy diffusion.

\subsubsection{Numerical simulations of shear driven turbulence in a stably stratified atmosphere with high thermal diffusivity}
\label{shearturb}

We  now present numerical simulations performed in an attempt to test Zahn's prescription for the radial turbulent transport in
differentially rotating radiative zones.
They have been performed in plan parallel geometry and with no effect of the Coriolis force.

Prat \& Lignières \cite{Prat2013,Prat2014} considered a flow configuration where a constant mean shear $d\langle U \rangle/dz$ and
a constant Brunt-Väisälä frequency are maintained by body forces.
The shear feeds a turbulent flow initialized by a statistically isotropic velocity field of given power spectral density.
The Richardson number can be tuned to reach a
statistically stationary state.
The Reynolds number of these simulations is high enough to get a $\propto k^{-5/3}$ power spectral 
density while the aspect ratio of the numerical domain ensures that between four and eight large structures in
each horizontal direction and at least six in the vertical direction
are present in the flow. The generic self-sustained mechanism of shear turbulence described by \cite{Waleffe1997} (see Sect.~\ref{shear}) is also observed 
in these simulations.

\begin{figure*}
\centering
\resizebox{\hsize}{!}{\includegraphics{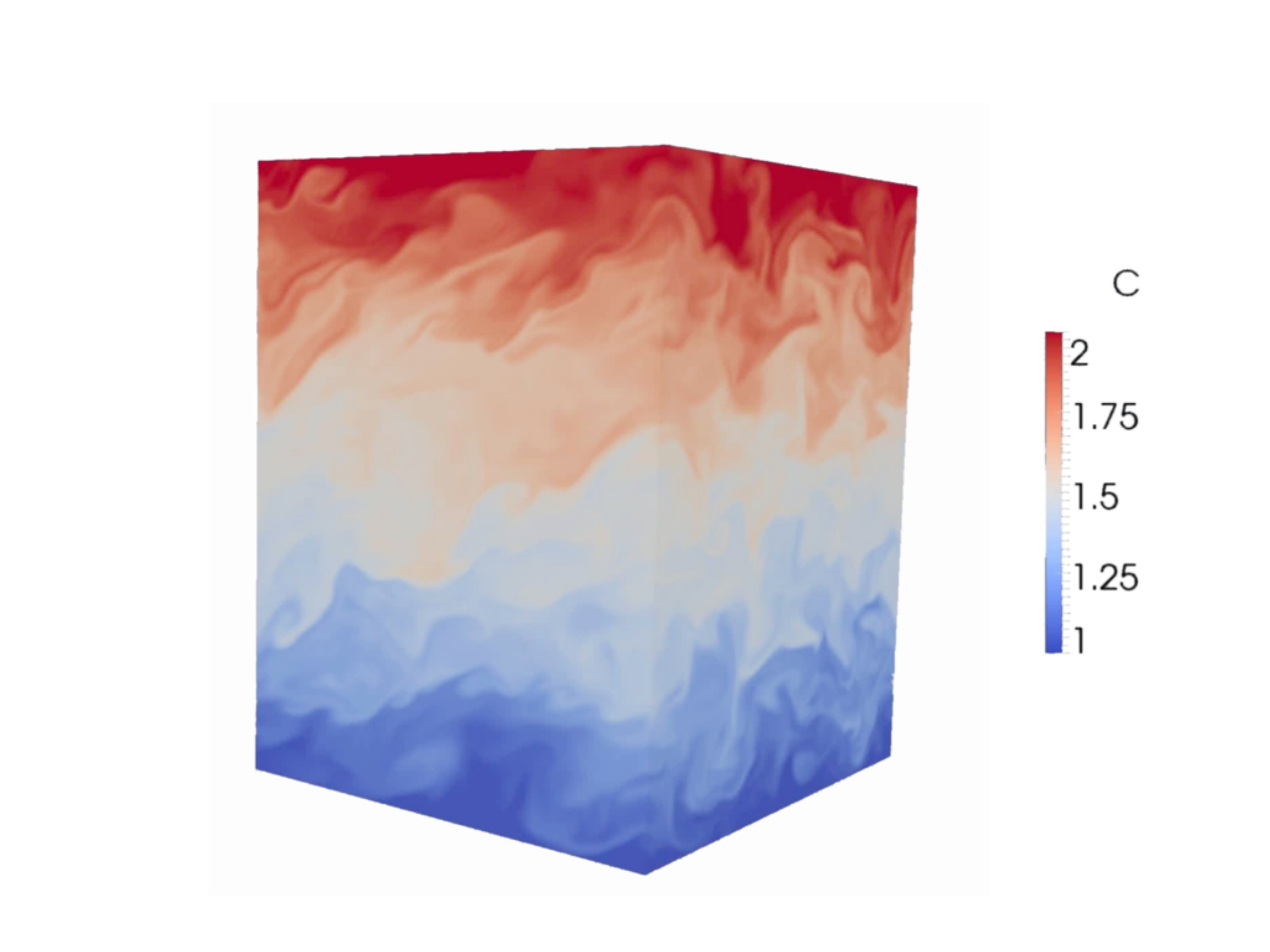}}
\caption{Snapshot of the concentration field when a mean vertical gradient $d\langle C\rangle/dz$ is imposed in a homogeneous
and stationary stably stratified shear turbulence \cite{Prat2013}.}
\label{Vincent}
\end{figure*}

These stationary and homogeneous turbulent shear flows are obtained on one hand using the Boussinesq equations with
decreasing values of the Prandtl number or equivalently of the Peclet number. In \cite{Prat2013,Prat2014} the Prandtl number is decreased from
$1$ to $10^{-3}$ while the Reynolds number remains fixed. This allowed us to reach a turbulent Peclet number of $0.34$ for a Reynolds number of $u \ell/\nu = 340$.
On the other hand,  
one small-Peclet-number simulation, obtained for a stationary $RiPe$, allows us to consider the asymptotic regime $Ri\gg 1, \; Pe \ll 1$ relevant for stars.

The vertical turbulent transport of a passive scalar is then determined 
using either the vertical dispersion of Lagrangian tracers,  the evolution
of the width of a Gaussian layer of passive scalar or the mean turbulent flux $\langle C'w' \rangle$ in the presence of a forced
gradient $d\langle C\rangle/dz$.  For illustration, Fig.~\ref{Vincent}
shows a snapshot of the concentration field in this last case. The two first methods show that 
the eddy diffusion model is approximatively valid in these simulations. The eddy diffusions obtained with the three methods are consistent. 

\begin{figure*}
\centering
\resizebox{\hsize}{!}{\includegraphics{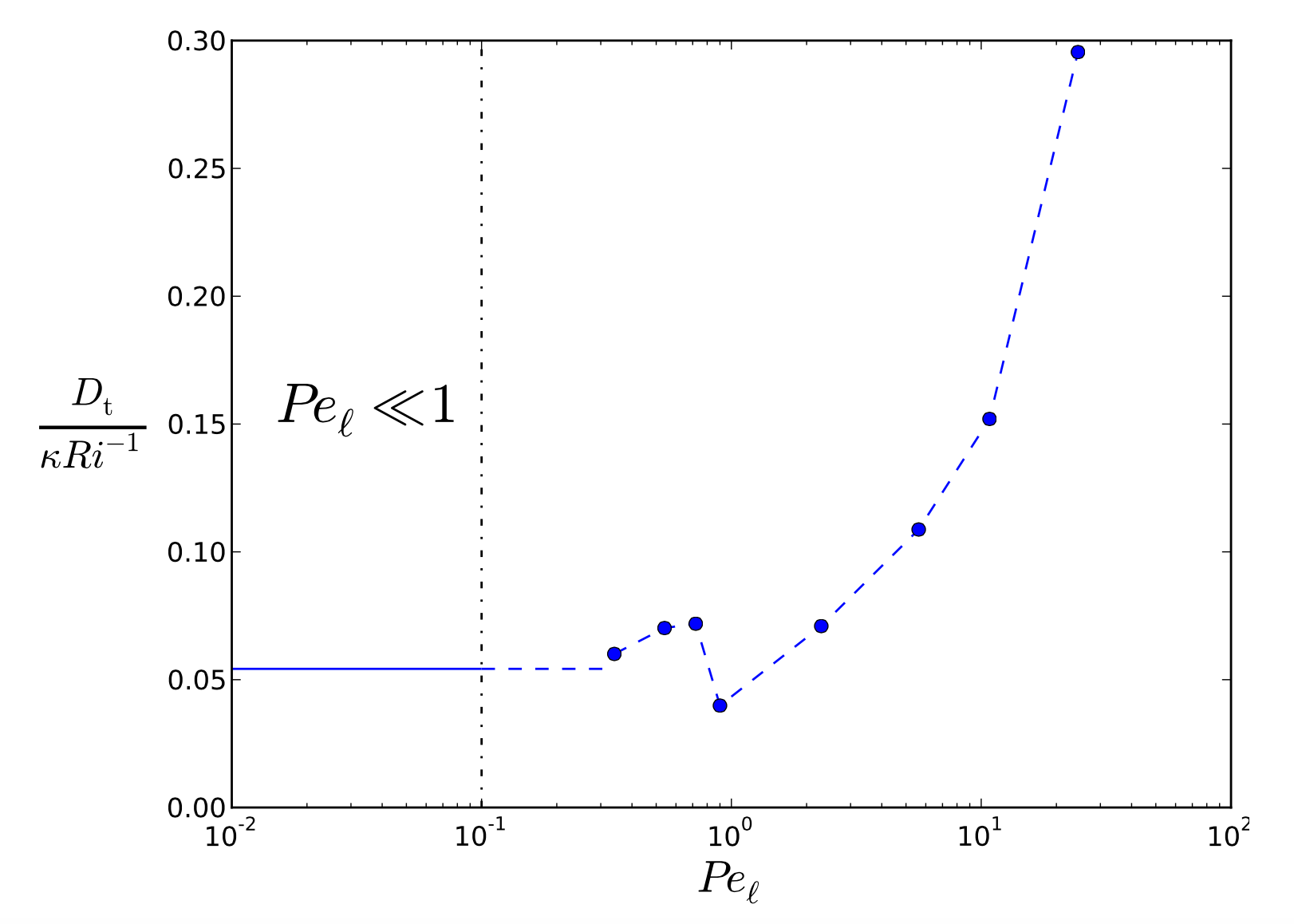}}
	\caption{$D_t/(\kappa Ri^{-1})$ as a function of the turbulent Peclet number. Dots
correspond to Boussinesq simulations. The solid line represents
the value obtained with the small-Peclet-number simulation.}
\label{Prat1}
\end{figure*}

The measured turbulent eddy diffusion scaled by the Zahn model is displayed in Fig.~\ref{Prat1} for the different numerical simulations
characterized by their turbulent Peclet numbers. They are shown for turbulent Peclet numbers 
around $1$ and for the asymptotic small-Peclet-approximation.
It appears that the Zahn scaling becomes valid for turbulent Peclet numbers of the order or smaller than one, and that
the asymptotic value of $D_t/(\kappa Ri^{-1})$ obtained with the small-Peclet-number equations is close to
the value found at the smallest turbulent Peclet number reached with the Boussinesq equations, $Pe_{\ell} = 0.34$.

\begin{figure*}
\centering
\resizebox{\hsize}{!}{\includegraphics{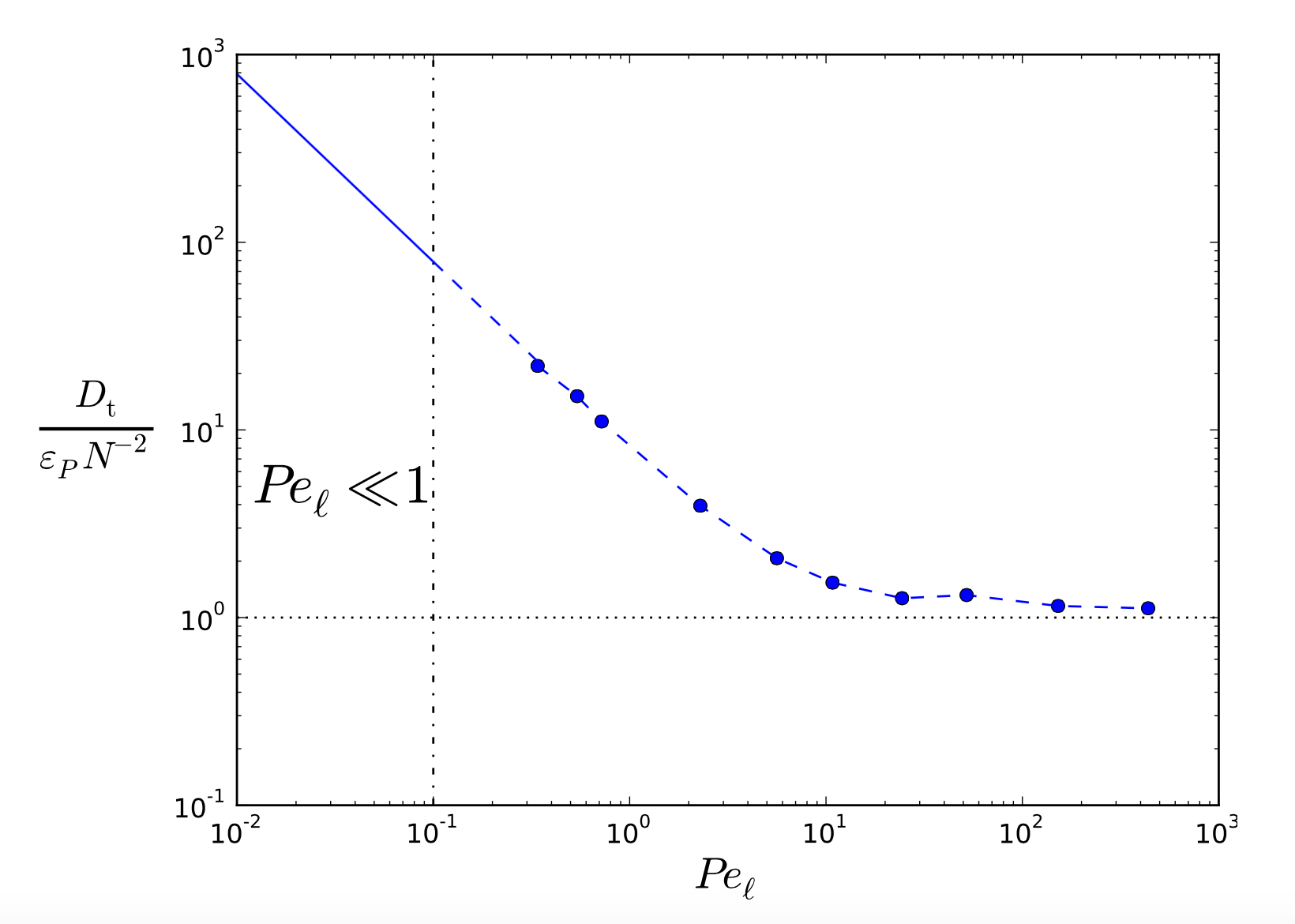}}
\caption{$D_t/(\epsilon_P N^{-2})$ as a function of the turbulent Peclet number. Dots
correspond to full Boussinesq simulations. The solid line represents
the value obtained with the small-Peclet-number simulation.}
\label{Prat2}
\end{figure*}

As expected, the Zahn scaling is not valid at higher Peclet number. In that regime, as shown in Fig.~\ref{Prat2},
the
eddy diffusivity $D_t = \epsilon_P/N^2$ fits the data much better.
We have seen in Sect.~\ref{examples} 
that this eddy diffusivity reproduces the vertical transport in the homogeneous stably stratified turbulence generated by a random forcing at $Pr\sim 1$.
In this regime, thermal diffusivity allows vertical transport by dissipating the buoyancy content of fluid parcels but this
process takes place at the dissipative length scales of turbulence in such a way that $\epsilon_P$, and thus $D_t$, does not depend on $\kappa$. 
This is no longer the case in the regime of small Peclet numbers where thermal diffusion plays a role in the large scale dynamics, and the vertical transport
linearly increases with $\kappa$. 

These results validate Zahn's scaling in the regime it was made for, that is when the transport is enabled by thermal diffusivity
in strongly stratified shear layer $Ri>1$. They also validate the small-Peclet-number approximation as a tool to explore
the asymptotic regime $Ri \gg 1, \;Pe \ll 1$. 

A series of papers \cite{Prat2016,Garaud2016, Garaud2017} have investigated the robustness of these results mostly by varying the Reynolds number
and the way the mean shear is forced in the simulations.
Most of them, except \cite{Garaud2016} who also performed Boussinesq simulations,
used the small-Peclet-number approximation for the parametric study.

In Prat et al. \cite{Prat2016}, it was found that $D_t/(\kappa Ri^{-1}) = 0.0356$ at $Re_{\ell}=7080$, the largest
Reynolds number considered in their study,
to be compared with $D_t/(\kappa Ri^{-1}) = 0.0558$ at $Re_{\ell}=336$ displayed in Fig.~\ref{Prat1}.
For these simulations, they used the shearing-box
configuration as a different way to force statistically constant shear and stratification, and found no difference
with the restoring forcing used in Prat \& Lignières \cite{Prat2013}.
Garaud et al. \cite{Garaud2017} considered a stratified plane Couette flow, where the shear is forced through the boundaries. 
The shear remains nevertheless quasi-uniform away from relatively small boundary layers. Their results are fully compatible
with those obtained by Prat et al. \cite{Prat2016}.

By decreasing the Reynolds number, both Prat et al. \cite{Prat2016} and Garaud et al. \cite{Garaud2017} found 
that $D_t/(\kappa Ri^{-1})$ increases slightly before it goes sharply to zero at $RiPr\sim 0.007$. In Sect.~\ref{lign99},
we already mentioned these results, as they provide the first determinations of the critical $RiPr_c$ for non-linear stability.
The dependence of $D_t/(\kappa Ri^{-1})$ at low Reynolds numbers, or equivalently at $RiPr$ smaller but close to $RiPr_c$, has 
been quantified through an empirical law in both studies.
Garaud et al. \cite{Garaud2017} in addition considered the regime of low $RiPe$ where the dynamics becomes unaffected by buoyancy.
Non-uniform shear profiles were also considered in two other numerical studies \cite{Garaud2016} and \cite{Gagnier2018}. In the latter, 
turbulence is sustained in
localized shear layers within the numerical domain which allows us to investigate the transport at the edge of the turbulent layer.

In short, for the regime $ Ri \gg 1$ relevant for stars, the Zahn model for transport and for stability 
has been confirmed in various numerical simulations that in turn provide in principle more reliable
values for $D_t/(\kappa R^{-1})$ and for $RiPr_c$. Despite some evidences given in \cite{Prat2016}, the assumption that $D_t/(\kappa R^{-1})$ 
will not vary by 
increasing the Reynolds number needs to be further confirmed. One must bear in mind that effective Reynolds numbers in stellar radiative zones
are not extremely high ($D_{\rm eff}/\nu \sim 20$ for the Lithium transport in the Sun \cite{Michaud1998} and about $1000-10000$ to account for
the angular momentum transport in the core of subgiants \cite{Eggenberger2012}).

Using the buoyancy modified time scale, Zahn's scaling can be derived as a mixing length model for shear flow, that is
$D_t \sim  \ell_m^2 d\langle U \rangle/dz$ where $\ell_m$ is specified by the condition that the shear time scale 
$1/(d\langle U\rangle/dz)$ equals the buoyancy modified time scale $t_{BM}$. 
This is a slightly more direct, although similar, derivation as compared to Zahn's that introduces a marginal stability argument.
The length scale that arises using this argument is $\ell_Z = \sqrt{(RiPe)_c} \ell_m$ and it has been named Zahn's scale in \cite{Garaud2017}.
The fact that the dynamically relevant lengthscale of the flow, $\ell_m$ or $\ell_Z$, can be directly related to the mean flow characteristics may be
the reason why such a simple parameterization of the turbulent transport exists.
This is not the case in turbulent stably stratified shear flows at $Pr\sim 1$ where numerical simulations of homogeneous shear turbulence
do not find an univocal relation between the eddy diffusivity and the Richardson number \cite{Shih2005}.

During stellar evolution, a stable gradient of chemical composition develops 
at the outer boundary of nuclear-burning
convective cores. This introduces an additional term in the buoyancy force that is not sensitive to thermal diffusion but rather to the much 
lower molecular diffusivity. The consequence is a potentially very efficient barrier for the radial transport,
with important effects on
the abundance of the CNO cycle elements at the surface of massive stars \cite{Meynet2013}.
Various prescriptions for the turbulent transport induced by a shear across the chemical composition gradient have
been proposed in the literature \cite{Maeder1996, Maeder1997, Talon1997}. 
They basically adapt the Zahn model to the presence of this additional stabilizing effect.
From their numerical simulations, Prat \& Lignières \cite{Prat2014} derived the following prescription :
\beqa
D_t &=& \alpha\kappa Ri^{-1}(Ri_{\sf cr} - Ri_\mu) \;\; \mathrm{with}\;\ \alpha =0.45\;\ \mathrm{and}\;\ Ri_{\sf cr} =0.12
\eeqa
\noindent where $Ri_\mu = N_\mu^2/(d\langle U\rangle/dz)^2$ is the Richardson number 
defined from the Brunt-Väisälä frequency $N_\mu^2 = - \beta' 
g d\langle \mu \rangle/dz$ associated with the gradient of the mean molecular weight $\mu$, $\beta'$  being the coefficient of compositional contraction
of the fluid.  
The model of \cite{Maeder1996} agrees with the numerical results while
the model proposed in \cite{Maeder1997} 
is not compatible
with them. However, according to \cite{Meynet2013}, the latter best fits the observations.
Thus, if we trust the results of numerical simulations, this comparison indicates the need for extra mixing 
at the edge of the massive star convective cores. 
In \cite{Talon1997}, the effect of an horizontal turbulence on diminishing the buoyancy content was proposed
as a possibility to increase the radial transport. The horizontal turbulence present in the homogeneous and stationary shear simulations
does not seem to have an effect but, in the context of \cite{Talon1997}, this turbulence is not attributed to the radial
 shear but rather to an additional horizontal shear. This point remains to be studied using
 numerical simulations.

Finally Prat et al. \cite{Prat2016} and Garaud et al. \cite{Garaud2017} also computed the turbulent flow of horizontal momentum 
$\langle u'w' \rangle$ in their simulations.
Deriving a turbulent viscosity defined by $\nu_t = \langle u'w' \rangle/d\langle U\rangle/dz$, \cite{Prat2016} found $\nu_t \sim 0.8 D_t$
when $RiPr < 3 \times 10^{-3}$ and in \cite{Garaud2017} $\nu_t/D_t$ was comprised between $0.8$ and $1$.
Note that determining a value of $\nu_t$ in this way provides an estimate of the horizontal momentum transport, but it
does not tell whether the gradient-diffusion approximation is correct or not.

\subsection{Vertical eddy diffusion in strongly stably stratified turbulence with high thermal diffusivity}
\label{strongturb}

In the previous section we considered the case of a dynamical radial shear instability that produces approximatively isotropic
motions at vertical length scales below $\ell_c= \sqrt{\kappa/N}$.
But other processes, other instabilities, may sustain turbulent motions of different types and we are obviously interested in their
transport properties. In Zahn (1992) \cite{Zahn1992}, the horizontal turbulence
that is invoked to strongly limit the latitudinal differential rotation is assumed to be generated by barotropic instabilities of 
the latitudinal differential rotation itself. Horizontal motions being authorized by stable stratification, this turbulence is 
believed to be predominantly horizontal  $w/u_h \sim \ell_v/\ell_h\ll 1$. This could also be the case for other sources of 
turbulence whose driving mechanism
is not constrained by the condition $\ell \leq \ell_c$ as it is the case for the dynamical instability of the vertical shear.
In this category one can think of the turbulent motions enforced by convective overshooting in the radiative zone
or the turbulence driven by baroclinic instabilities \cite{Spruit1984}.

As already mentioned the atmospheric and oceanic largescale motions are characterized by very low Froude numbers $Fr_h = u_h/(N \ell_h)$ (see Tab.~\ref{table1}).
This fact has motivated studies where a turbulent flow is forced at low horizontal Froude number, this situation being referred to as strongly stratified turbulence.
A theoretical approach of strongly stratified turbulence is to consider the asymptotic limit of the Boussinesq equation in the $Fr \to 0$ limit.
In the scaling analysis of Riley et al. \cite{Riley1981} and Lilly \cite{Lilly1983}, the vertical Froude number defined by $Fr_v = u_h/(N \ell_v)$ is assumed to vanish also when $Fr_h \to 0$, 
leading to asymptotic equations
where the horizontal flow is governed by purely two-dimensional equations. 
Billant \& Chomaz \cite{Billant2001} reconsidered this scaling without a priori assumption on the vertical length $\ell_v$. They built on
the self-similar properties of the invisicd $Fr_h \ll 1$ equations with respect to $zN/u_h$ to show that 
the vertical length scale adapts to the system
through $\ell_v \sim u_h/N$, that is $Fr_v \sim 1$.
The resulting asymptotic equations are not two-dimensional, in particular ruling out the possibility of an inverse cascade.
They instead predict a direct energy cascade with an horizontal energy spectrum $E_h(k_h) \sim \epsilon^{2/3} k_h^{-5/3}$.
Various numerical simulations have confirmed the relevance of this scaling analysis in the $Fr_h \ll 1$ regime,
as long as the buoyancy Reynolds number is high enough  (${\cal R} = \frac{\epsilon}{N^2\nu} \gtrsim 10$)
\cite{Bretou2007, Waite2011,Bartello2013,Augier2015, Maffioli2016}.
 
Thus, in the double limit
$Fr_h \ll 1, {\cal R} \gg 1$, we may assume that
turbulence is strongly anisotropic with $\ell_v/\ell_h \sim w/u_h \sim Fr_h$.
Horizontal motions show a direct energy cascade
in between $\ell_h$ and the Ozmidov scale $\ell_O = (\frac{\epsilon}{N^3})^{1/2}$ that separates the lengthscales affected by buoyancy from
the isotropic inertial range. 
Below $\ell_O$, the return to isotropy is possible
and a Kolmogorov turbulent cascade develops between $\ell_O$ and the Kolmogrov scale $ \eta = (\nu^3/\epsilon)^{1/4}$. The condition
${\cal R} \gg 1$ is crucial for the existence of the two inertial ranges below and above $\ell_O$ \cite{Bretou2007}.

An extrapolation of these results to stellar conditions can be attempted as follows. Two regimes are to be distinguished depending on the
rate of energy input into the turbulence  $u_h^3/\ell_h \sim \epsilon$.
If it is high enough so that return to isotropy takes place at lengthscales which are not affected by thermal diffusion, the previous scaling holds.
On the other hand, if $\ell_c = \sqrt{\kappa/N} > \ell_O$, the buoyancy effects will be strongly diminished at scales larger than $\ell_O$.
In this case, we can define a modified Ozmidov scale $\ell_{OM} = \left( \kappa \epsilon^{1/3}/N^2)\right)^{3/8}$for return to isotropy as
the scale for which the eddy turnover time $\ell/u$, where $u=(\epsilon \ell)^{1/3}$, is equal to
the modified buoyancy time $\kappa/(N^2 \ell^2)$. 

In Sect.~(\ref{modelling}), we have seen that when $Pr\sim 1$ 
the vertical eddy diffusivity in strongly stratified turbulence reads  $D_t = \Gamma \epsilon/N^2$ with $\Gamma \sim 0.33$ in the
$Fr_h \to 0$ limit. This turbulent diffusivity can be equivalently expressed using the Ozmidov scale as $D_t = \Gamma u_O \ell_O$ with $u_O=(\epsilon \ell_O)^{1/3}$.
Using the same expression but with the modified Ozimdov scale yields an estimate of the vertical transport in the
$\ell_c > \ell_O$ regime. Finally, for strongly stratified turbulence in radiative zones, two regimes of vertical 
turbulent transport are expected depending on the characteristics of the horizontal turbulence, namely :
\beqa 	
\mathrm{if}\;\;\ell_O > \ell_c & \;\; \rightarrow &  \;\; D_t = \Gamma \frac{\epsilon}{N^2} \\
\mathrm{if}\;\;\ell_c > \ell_O & \;\; \rightarrow & \;\; D_t = \Gamma' \left(\frac{\epsilon \kappa}{N^2}\right)^{1/2}
\eeqa
\noindent where $\Gamma'$ is a yet unknown constant that plays the same role as $\Gamma$ .
This derivation of a vertical eddy diffusion in a strongly stratified turbulence $Fr_h \ll 1$ affected by thermal diffusion has not 
been published elsewhere.
Nevertheless, I found the same expression in \cite{Zahn1992} as a model for the vertical turbulence driven by horizontal shears.
According to \cite{Zahn1992}, it has been derived in \cite{Schatzman1991} following Townsend's arguments \cite{Townsend1958} to account for thermal diffusion effects.

It turns out that a quite general statement can be deduced by combining these estimates and the observational constraints showing 
that the  effective transport of chemicals
is in general much smaller than the thermal diffusivity, $D_{\rm eff} \ll \kappa$ (see Sect.~\ref{intro}).
Indeed, a consequence of the previous scalings is that $D_t > \kappa$ when $\ell_O > \ell_c$, whereas $D_t < \kappa$ when $\ell_c > \ell_O$.
Thus we conclude that if turbulent motions are indeed responsible for the observed transport, the fact that $D_{\rm eff} < \kappa$ implies
that the eddies involved in the vertical transport 
are affected by thermal diffusion, that is their length scales are smaller than  $\ell_c =\sqrt{\kappa/N}$.
This same conclusion was reached by \cite{Lignieres2005} using similar arguments regarding thermal diffusion effects, but
without reference to the 
strongly stratified turbulence scalings of \cite{Bretou2007}.

\section{Conclusion}
\label{conc}

We have reviewed physical processes involved in the vertical turbulent transport produced by a plan parallel shear flow in a vertically stably stratified atmosphere
with high thermal diffusivity. This led us to discuss shear instabilities, eddy diffusion, and stratified turbulence in 
conditions encountered in the Earth fluid envelope ($Pr \sim 1$) up to stellar radiative zones  ($Pr \ll 1$).
We showed that numerical simulations have been successful in testing Zahn's model for the radial turbulent transport induced by radial differential rotation.

Regarding the broader issue of modeling the transport of chemical elements in differentially rotating radiative zones, many more questions can be approached through
dedicated numerical simulations. Local simulations such as the ones presented here are well suited to investigate turbulent transport processes
occurring at small length scales. But they do not provide informations on the largescale flow. This requires global simulations
in spherical geometry. As a first step, this type of simulation should tell us more about the differential rotation and the laminar largescale flows
driven when torques are applied to radiative zones. These torques might be due to stellar winds, or to
Reynolds stresses at the interface with a differentially convective zone or to structural changes during expansion or contraction phases.
A better knowledge of the resulting large scale configurations should help us identify the instabilities than can power a transition to turbulence,
and provide us with some constraints on the scale of the dominant eddies. For example the main instabilities 
triggered in a differentially rotating star embedded in a poloidal magnetic field are being studied 
thanks to 
a combination of axisymmetric 
and 3D simulations \cite{Jouve2015, Gaurat2015}.
In non-magnetic radiative zones, global numerical simulations should be designed to test the hypothesis of weak latitudinal differential rotation
which is at the base of the current models of rotationally induced transport.

\bibliography{hdr.bib}

\begin{thebibliography}{90}

\bibitem{Michaud04}
G.~{Michaud}, \emph{{Atomic diffusion in stellar surfaces and interiors}}, in
  \emph{The A-Star Puzzle, IAU Symposium 224}, edited by {J.~Zverko,
  J.~Ziznovsky, S.~J.~Adelman, \& W.~W.~Weiss} (Cambridge University Press,
  2004), pp. 173--183

\bibitem{Hunter2009}
I.~{Hunter}, I.~{Brott}, N.~{Langer}, D.J. {Lennon}, P.L. {Dufton}, I.D.
  {Howarth}, R.S.I. {Ryans}, C.~{Trundle}, C.J. {Evans}, A.~{de Koter} et~al.,
  \aap \textbf{496}, 841 (2009), \texttt{0901.3853}

\bibitem{Pinsonneault1997}
M.~{Pinsonneault}, \araa \textbf{35}, 557 (1997)

\bibitem{Thompson1996}
M.J. {Thompson}, J.~{Toomre}, E.R. {Anderson}, H.M. {Antia}, G.~{Berthomieu},
  D.~{Burtonclay}, S.M. {Chitre}, J.~{Christensen-Dalsgaard}, T.~{Corbard},
  M.~{De Rosa} et~al., Science \textbf{272}, 1300 (1996)

\bibitem{Mosser2012}
B.~{Mosser}, M.J. {Goupil}, K.~{Belkacem}, J.P. {Marques}, P.G. {Beck},
  S.~{Bloemen}, J.~{De Ridder}, C.~{Barban}, S.~{Deheuvels}, Y.~{Elsworth},
  \aap \textbf{548}, A10 (2012), \texttt{1209.3336}

\bibitem{Deheuvels2014}
S.~{Deheuvels}, G.~{Do{\u{g}}an}, M.J. {Goupil}, T.~{Appourchaux},
  O.~{Benomar}, H.~{Bruntt}, T.L. {Campante}, L.~{Casagrande}, T.~{Ceillier},
  G.R. {Davies}, \aap \textbf{564}, A27 (2014), \texttt{1401.3096}

\bibitem{VanReeth2016}
T.~{Van Reeth}, A.~{Tkachenko}, C.~{Aerts}, \aap \textbf{593}, A120 (2016),
  \texttt{1607.00820}

\bibitem{Salaris2017}
M.~{Salaris}, S.~{Cassisi}, Royal Society Open Science \textbf{4}, 170192
  (2017), \texttt{1707.07454}

\bibitem{Michaud1998}
G.~{Michaud}, J.P. {Zahn}, Theoretical and Computational Fluid Dynamics
  \textbf{11}, 183 (1998)

\bibitem{Korn2006}
A.J. {Korn}, F.~{Grundahl}, O.~{Richard}, P.S. {Barklem}, L.~{Mashonkina},
  R.~{Collet}, N.~{Piskunov}, B.~{Gustafsson}, Nature \textbf{442}, 657 (2006),
  \texttt{astro-ph/0608201}

\bibitem{Eggenberger2012}
P.~{Eggenberger}, J.~{Montalb{\'a}n}, A.~{Miglio}, \aap \textbf{544}, L4
  (2012), \texttt{1207.1023}

\bibitem{Zahn1992}
J.~{Zahn}, \aap \textbf{265}, 115 (1992)

\bibitem{Meynet2000}
G.~{Meynet}, A.~{Maeder}, \aap \textbf{361}, 101 (2000),
  \texttt{astro-ph/0006404}

\bibitem{Maeder2001}
A.~{Maeder}, G.~{Meynet}, \aap \textbf{373}, 555 (2001),
  \texttt{astro-ph/0105051}

\bibitem{Talon2005}
S.~{Talon}, C.~{Charbonnel}, \aap \textbf{440}, 981 (2005),
  \texttt{astro-ph/0505229}

\bibitem{Zahn2013}
J.P. {Zahn}, \emph{{Rotation induced mixing in stellar interiors}}, in
  \emph{EAS Publications Series}, edited by G.~{Alecian}, Y.~{Lebreton},
  O.~{Richard}, G.~{Vauclair} (2013), Vol.~63, pp. 245--254

\bibitem{Spruit2002}
H.C. {Spruit}, \aap \textbf{381}, 923 (2002), \texttt{arXiv:astro-ph/0108207}

\bibitem{Thorpe1971}
S.A. {Thorpe}, Journal of Fluid Mechanics \textbf{46}, 299 (1971)

\bibitem{Batch1967}
G.K. {Batchelor}, \emph{{An Introduction to Fluid Dynamics}} (Cambridge
  University Press, 1967)

\bibitem{Baines1994}
P.G. {Baines}, H.~{Mitsudera}, Journal of Fluid Mechanics \textbf{276}, 327
  (1994)

\bibitem{Vallis2006}
G.K. {Vallis}, \emph{{Atmospheric and Oceanic Fluid Dynamics}} (Cambridge
  University Press, 2006)

\bibitem{Caulfield1995}
C.P. {Caulfield}, W.R. {Peltier}, S.~{Yoshida}, M.~{Ohtani}, Physics of Fluids
  \textbf{7}, 3028 (1995)

\bibitem{Drazin2004}
P.G. Drazin, W.H. Reid, \emph{Hydrodynamic Stability}, Cambridge Mathematical
  Library, 2nd~edn. (Cambridge University Press, 2004)

\bibitem{Grossmann2000}
S.~{Grossmann}, Reviews of Modern Physics \textbf{72}, 603 (2000)

\bibitem{Schmid2007}
P.J. {Schmid}, Annual Review of Fluid Mechanics \textbf{39}, 129 (2007)

\bibitem{Schmid2012}
P.~Schmid, D.~Henningson, \emph{{Stability and Transition in Shear Flows}},
  Applied Mathematical Sciences (Springer New York, 2012)

\bibitem{Trefethen1993}
L.N. {Trefethen}, A.E. {Trefethen}, S.C. {Reddy}, T.A. {Driscoll}, Science
  \textbf{261}, 578 (1993)

\bibitem{Hamilton1995}
J.M. {Hamilton}, J.~{Kim}, F.~{Waleffe}, Journal of Fluid Mechanics
  \textbf{287}, 317 (1995)

\bibitem{Waleffe1997}
F.~{Waleffe}, Physics of Fluids \textbf{9}, 883 (1997)

\bibitem{Eckhardt2018}
B.~{Eckhardt}, Physica A Statistical Mechanics and its Applications
  \textbf{504}, 121 (2018), \texttt{1801.09260}

\bibitem{Rincon2007}
F.~{Rincon}, G.I. {Ogilvie}, M.R.E. {Proctor}, \prl \textbf{98}, 254502 (2007),
  \texttt{0705.2814}

\bibitem{Zahn1974}
J.P. {Zahn}, \emph{{Rotational Instabilities and Stellar Evolution}}, in
  \emph{Stellar Instability and Evolution}, edited by P.~{Ledoux}, A.~{Noels},
  A.W. {Rodgers} (1974), Vol.~59 of \emph{IAU Symposium}, p. 185

\bibitem{Chandra1961}
S.~Chandrasekhar, \emph{Hydrodynamic and Hydromagnetic Stability},
  International series of monographs on physics (Clarendon Press, 1961)

\bibitem{Miles1986}
J.~{Miles}, Physics of Fluids \textbf{29}, 3470 (1986)

\bibitem{Rieutord2015}
M.~Rieutord, \emph{{Fluid Dynamics: An Introduction}}, Graduate Texts in
  Physics (Springer International Publishing, 2015)

\bibitem{Spiegel1960}
E.A. {Spiegel}, G.~{Veronis}, \apj \textbf{131}, 442 (1960)

\bibitem{Drazin1958}
P.G. {Drazin}, Journal of Fluid Mechanics \textbf{4}, 214 (1958)

\bibitem{Smyth1988}
W.D. {Smyth}, G.P. {Klaassen}, W.R. {Peltier}, Geophysical and Astrophysical
  Fluid Dynamics \textbf{43}, 181 (1988)

\bibitem{Peltier2003}
W.R. {Peltier}, C.~{Caulfield}, Annual Review of Fluid Mechanics \textbf{35},
  135 (2003)

\bibitem{Witzke2015}
V.~{Witzke}, L.J. {Silvers}, B.~{Favier}, \aap \textbf{577}, A76 (2015),
  \texttt{1503.02790}

\bibitem{Talon2008}
S.~{Talon}, C.~{Charbonnel}, \aap \textbf{482}, 597 (2008), \texttt{0801.4643}

\bibitem{Lign99}
F.~{Ligni{\`e}res}, F.~{Califano}, A.~{Mangeney}, \aap \textbf{349}, 1027
  (1999), \texttt{arXiv:astro-ph/9908184}

\bibitem{Dudis1974}
J.J. {Dudis}, Journal of Fluid Mechanics \textbf{64}, 65 (1974)

\bibitem{Jones1977}
C.A. {Jones}, Geophysical and Astrophysical Fluid Dynamics \textbf{8}, 165
  (1977)

\bibitem{Goldreich1967}
P.~{Goldreich}, G.~{Schubert}, \apj \textbf{150}, 571 (1967)

\bibitem{Townsend1958}
A.A. {Townsend}, Journal of Fluid Mechanics \textbf{4}, 361 (1958)

\bibitem{Prat2016}
V.~{Prat}, J.~{Guilet}, M.~{Viallet}, E.~{M{\"u}ller}, \aap \textbf{592}, A59
  (2016), \texttt{1512.04223}

\bibitem{Garaud2017}
P.~{Garaud}, D.~{Gagnier}, J.~{Verhoeven}, \apj \textbf{837}, 133 (2017),
  \texttt{1610.04320}

\bibitem{Taylor1921}
G.I. {Taylor}, Proc. London Math. Soc. \textbf{20}, 196–211 (1921)

\bibitem{Richardson1926}
L.F. {Richardson}, Proceedings of the Royal Society of London Series A
  \textbf{110}, 709 (1926)

\bibitem{Sawford2001}
B.~{Sawford}, Annual Review of Fluid Mechanics \textbf{33}, 289 (2001)

\bibitem{Prandtl1925}
L.~{Prandtl}, Zeitschrift Angewandte Mathematik und Mechanik \textbf{5}, 136
  (1925)

\bibitem{Tennekes}
H.~Tennekes, J.~Lumley, \emph{A First Course in Turbulence} (MIT Press, 1972)

\bibitem{Gregg2018}
M.C. {Gregg}, E.A. {D'Asaro}, J.J. {Riley}, E.~{Kunze}, Annual Review of Marine
  Science \textbf{10}, 443 (2018)

\bibitem{Ledwell1993}
J.R. {Ledwell}, A.J. {Watson}, C.S. {Law}, \nat \textbf{364}, 701 (1993)

\bibitem{Ledwell2000}
J.R. {Ledwell}, E.T. {Montgomery}, K.L. {Polzin}, L.C. {St. Laurent}, R.W.
  {Schmitt}, J.M. {Toole}, \nat \textbf{403}, 179 (2000)

\bibitem{Lindborg2009}
E.~{Lindborg}, E.~{Fedina}, Geophysics Research Letters \textbf{36}, L01605
  (2009)

\bibitem{van2008}
M.~{van Aartrijk}, H.J.H. {Clercx}, K.B. {Winters}, Physics of Fluids
  \textbf{20}, 025104-025104-16 (2008)

\bibitem{Kaneda2000}
Y.~{Kaneda}, T.~{Ishida}, Journal of Fluid Mechanics \textbf{402}, 311 (2000)

\bibitem{Pearson1983}
H.J. {Pearson}, J.S. {Puttock}, J.C.R. {Hunt}, Journal of Fluid Mechanics
  \textbf{129}, 219 (1983)

\bibitem{Lindborg2008}
E.~{Lindborg}, G.~{Brethouwer}, Journal of Fluid Mechanics \textbf{614}, 303
  (2008)

\bibitem{Bretou2009}
G.~{Brethouwer}, E.~{Lindborg}, Journal of Fluid Mechanics \textbf{631}, 149
  (2009)

\bibitem{Osborn1972}
T.R. {Osborn}, C.S. {Cox}, Geophysical and Astrophysical Fluid Dynamics
  \textbf{3}, 321 (1972)

\bibitem{Osborn1980}
T.R. {Osborn}, Journal of Physical Oceanography \textbf{10}, 83 (1980)

\bibitem{Maffioli2016}
A.~{Maffioli}, P.A. {Davidson}, Journal of Fluid Mechanics \textbf{786}, 210
  (2016)

\bibitem{Qi2017}
Q.~{Zhou}, J.R. {Taylor}, C.P. {Caulfield}, Journal of Fluid Mechanics
  \textbf{820}, 86 (2017)

\bibitem{LignPe}
F.~{Ligni{\`e}res}, \aap \textbf{348}, 933 (1999),
  \texttt{arXiv:astro-ph/9908182}

\bibitem{Donatelli2016}
D.~{Donatelli}, B.~{Ducomet}, M.~{Kobera}, S.~{Necasova}, Electronic Journal of
  Differential Equations \textbf{245}, 1 (2016)

\bibitem{Gough1969}
D.O. {Gough}, Journal of Atmospheric Sciences \textbf{26}, 448 (1969)

\bibitem{Prat2013}
V.~{Prat}, F.~{Ligni{\`e}res}, \aap \textbf{551}, L3 (2013), \texttt{1301.4151}

\bibitem{Garaud2016}
P.~{Garaud}, L.~{Kulenthirarajah}, \apj \textbf{821}, 49 (2016),
  \texttt{1512.08774}

\bibitem{Prat2014}
V.~{Prat}, F.~{Ligni{\`e}res}, \aap \textbf{566}, A110 (2014),
  \texttt{1404.6199}

\bibitem{Gagnier2018}
D.~{Gagnier}, P.~{Garaud}, \apj \textbf{862}, 36 (2018), \texttt{1803.10455}

\bibitem{Shih2005}
L.H. {Shih}, J.R. {Koseff}, G.N. {Ivey}, J.H. {Ferziger}, Journal of Fluid
  Mechanics \textbf{525}, 193 (2005)

\bibitem{Meynet2013}
G.~{Meynet}, S.~{Ekstrom}, A.~{Maeder}, P.~{Eggenberger}, H.~{Saio},
  V.~{Chomienne}, L.~{Haemmerl{\'e}}, \emph{{Models of Rotating Massive Stars:
  Impacts of Various Prescriptions}}, in \emph{Studying Stellar Rotation and
  Convection}, edited by {M.J.~Goupil, K.~Belkacem, C.~Neiner, F.
  Ligni{\`e}res, \& J.J.~Green} (Lecture Notes in Physics, Springer-Verlag
  Berlin Heidelberg, 2013), Vol. 865, p.~3

\bibitem{Maeder1996}
A.~{Maeder}, G.~{Meynet}, \aap \textbf{313}, 140 (1996)

\bibitem{Maeder1997}
A.~{Maeder}, \aap \textbf{321}, 134 (1997)

\bibitem{Talon1997}
S.~{Talon}, J.P. {Zahn}, \aap \textbf{317}, 749 (1997),
  \texttt{astro-ph/9609010}

\bibitem{Spruit1984}
H.C. {Spruit}, E.~{Knobloch}, \aap \textbf{132}, 89 (1984)

\bibitem{Riley1981}
J.J. {Riley}, R.W. {Metcalfe}, M.A. {Weissman}, \emph{{Direct numerical
  simulations of homogeneous turbulence in density-stratified fluids}}, in
  \emph{American Institute of Physics Conference Series} (1981), Vol.~76, pp.
  79--112

\bibitem{Lilly1983}
D.K. {Lilly}, Journal of Atmospheric Sciences \textbf{40}, 749 (1983)

\bibitem{Billant2001}
P.~{Billant}, J.M. {Chomaz}, Physics of Fluids \textbf{13}, 1645 (2001)

\bibitem{Bretou2007}
G.~{Brethouwer}, P.~{Billant}, E.~{Lindborg}, J.M. {Chomaz}, Journal of Fluid
  Mechanics \textbf{585}, 343 (2007)

\bibitem{Waite2011}
M.L. {Waite}, Physics of Fluids \textbf{23}, 066602-066602-12 (2011)

\bibitem{Bartello2013}
P.~{Bartello}, S.M. {Tobias}, Journal of Fluid Mechanics \textbf{725}, 1 (2013)

\bibitem{Augier2015}
P.~{Augier}, P.~{Billant}, J.M. {Chomaz}, Journal of Fluid Mechanics
  \textbf{769}, 403 (2015)

\bibitem{Schatzman1991}
E.~{Schatzman}, A.~{Baglin}, \aap \textbf{249}, 125 (1991)

\bibitem{Lignieres2005}
F.~{Ligni{\`e}res}, N.~{Toqu{\'e}}, A.~{Vincent}, \emph{{Turbulent transport in
  stellar radiative zones}}, in \emph{Elements Stratification in Stars: 40
  years of Atomic Diffusion}, edited by {G.~Alecian, O.~Richard, \&
  S.~Vauclair} (EAS Publications Series, EDP Sciences, 2005), pp. 209--214

\bibitem{Jouve2015}
L.~{Jouve}, T.~{Gastine}, F.~{Ligni{\`e}res}, \aap \textbf{575}, A106 (2015),
  \texttt{1412.2900}

\bibitem{Gaurat2015}
M.~{Gaurat}, L.~{Jouve}, F.~{Ligni{\`e}res}, T.~{Gastine}, \aap \textbf{580},
  A103 (2015), \texttt{1507.01508}

\end{thebibliography}

\end{document}